\documentclass[aps,prd,reprint,longbibliography]{revtex4-2}
\usepackage[utf8]{inputenc}
\usepackage{graphicx,fancybox,float,comment}

\usepackage{units}
\usepackage{multirow,array}
\usepackage[dvipsnames]{xcolor}
\usepackage{amsmath,amssymb,amsfonts,amsthm}

\theoremstyle{definition}
\newtheorem{definition}{Definition}[section]

\usepackage{yfonts}
\usepackage{tensor}
\usepackage{paralist}
\usepackage{braket}
\usepackage{tensor,mathrsfs}
\usepackage[normalem]{ulem}
\usepackage{pifont}
\usepackage{hyperref}

\usepackage{cleveref}

\begin{document}
\title{An example of the convergence of hydrodynamics in strong external fields}

\newcommand{\uuAff}{\affiliation{Institute for Theoretical Physics, Utrecht University, Princetonplein 5, 3584 CC Utrecht, The Netherlands}}
\author{Casey Cartwright}
\email{c.c.cartwright@uu.nl}
\uuAff
\date{\today}

\begin{abstract}
The anti-de-Sitter/conformal field theory (AdS/CFT) correspondence is used to provide an estimate of the radius of convergence of the linearized gradient expansion of the hydrodynamic description of $\mathcal{N}=4$ supersymmetric Yang-Mills (SYM) theory minimally coupled to $U(1)$ gauge theory subjected to strong magnetic fields. The results of this work demonstrate that the dispersion relations of hydrodynamic modes continue to converge for magnetic field strengths far beyond those values for which a hydrodynamic description is expected. For magnetic field strengths much larger then the temperature scale the bulk dual interpolates between AdS$_{4+1}$ and the product of a Ba\~{n}ados-Teitelboim-Zanelli (BTZ) black hole and a two dimensional manifold (BTZ$_{2+1}\times \mathbb{R}^2$), and may be regarded as a renormalization group flow of the $3+1$ dimensional CFT to a $1+1$ dimensional CFT. In this regime, we clarify past literature on the quasi-normal mode (QNM) spectrum of bulk scalar fields by introducing a new way to classify the behavior of QNM collisions in the complex frequency and momentum plane.
\end{abstract}

\newcommand{\exd }{\mathrm{d}}
\newcommand{\tr}{\mathrm{Tr}}
\newcommand{\q}{\mathfrak{q}}
\newcommand{\w}{\mathfrak{w}}
\maketitle

%%%%%%%%%%%%%%%%%%%%%
\section{Introduction}
\label{sec:intro}
In general, hydrodynamics can be described as, a universal, effective field theory, of the near equilibrium collective behavior of many body (quantum) systems described by the evolution of (non)-conserved quantities. In the case of quantum many body system without any internal symmetries the only relevant conserved current is the energy momentum tensor associated with the thermal degrees of freedom. If the underlying microscopic theory is weakly coupled to $U(1)$ gauge fields there are varying degrees of complexity at which one can treat the system depending on the coupling between the thermal and electromagnetic degrees of freedom.

If the coupling between thermal and electromagnetic degrees of freedom is negligible one can separate the two sectors and separately consider the hydrodynamic approximation of the thermal degrees of freedom and solve Maxwell’s equations in matter~\cite{Grozdanov:2016tdf,Hernandez:2017mch}
\begin{align}
\nabla\cdot D &= \rho_f \, , \quad \nabla\cdot B=0\, , \\
\nabla \times H &=J_f+\partial_t D\, , \quad \nabla \times E=\partial_t B \, .
\end{align}
However, if the coupling between thermal and electromagnetic degrees of freedom is non-negligible, one is required to solve hydrodynamics coupled to Maxwell's equation in matter. This comes in two distinct flavors depending on whether you consider the fields that comprise the gauge field to be external, or dynamic. If one treats, the fields as external, then relevant hydrodynamic variables are $u^{\mu}, \, T , \, \mu$ and the equations of motion are
\begin{align}
\nabla_{\mu}T^{\mu\nu}&=F^{\nu\lambda}J_\lambda \\
\nabla_{\mu} J^{\mu}&=0
\end{align}
Otherwise if one treats, the fields as dynamical, then relevant hydrodynamic variables are  $u^{\mu}, \, T , \, \mu, \,E^{\mu}, \, B^{\mu}$ with the equations of motion~\cite{Grozdanov:2016tdf,Hernandez:2017mch}
\begin{align}
\nabla_{\mu}T^{\mu\nu}&=F^{\nu\lambda}J^{ext}_\lambda \, , \\
\nabla_{\nu} \left(F^{\mu\nu}-M^{\mu\nu} \right)&=J^{\mu}_{ext}+J^{\mu}_{free}\, , \\
\epsilon^{\mu\nu\alpha\beta} \nabla_\nu F_{\alpha\beta}&=0 \, .
\end{align}
In both scenarios it is relevant to consider the order of the magnetic field, in particular in both cases it is possible to have strong fields,
\begin{equation}
    E\sim O(\partial) \hspace{0.3cm} B\sim O(1)\, .
\end{equation}
In the first case this defines:
\noindent \textit{Strong field external hydrodyanmics} – The fields are external, the coupling between thermal and electromagnetic fields is non-negligible and the matter is electrically conducting while the electric field is screened.

In the second case it defines:
\noindent \textit{Magnetohydrodynamics (MHD)} – The fields are dynamical, the coupling between thermal and electromagnetic fields is non-negligible and the matter is electrically conducting while the electric field is screened.

In addition to the breakdown of the coupling between the thermal and electromagnetic degrees of freedom one can also consider systems whose microscopic theory contains an anomalous symmetry. Here, one must adjust the equations of motion for hydrodyanmics to include sources on the right hand side of the equations to encode the non-conservation induced by the anomaly. Indeed, this is precisely the case if one would like to use hydrodynamics as a model of the chiral magnetic effect (CME).

The CME describes the macroscopic charged current flow along a magnetic field in a theory with a chiral anomaly~\cite{Kharzeev:2004ey,Kharzeev:2007jp,Son:2009tf,Gynther:2010ed,Amado:2011zx,Kharzeev:2012ph,Ammon:2020rvg,Ammon:2017ded}. While successfully detected in condensed matter experiments (see for instance~\cite{Li:2014bha,Arnold:2015vvs,Huang:2015,Xiong:2015,Li:2015,Li:2016,Hirschberger:2016,Gooth:2017mbd,Shekhar:2018}) this effect has yet to be conclusively discovered in collider experiments (see for instance~\cite{STAR:2009tro,STAR:2009wot,STAR:2013ksd,STAR:2013zgu,STAR:2014uiw,STAR:2019xzd,STAR:2021mii} from the STAR collaboration or~\cite{ALICE:2012nhw,ALICE:2017sss,ALICE:2020siw,CMS:2016wfo,CMS:2017lrw} from ALICE and CMS), specifically the RHIC-BES. Hydrodynamic modeling is essential to interpret data collected from these highly energetic collisions. However, as is needed for the chiral magnetic effect, large magnetic fields are generated by the colliding nuclei. Hence in principle chiral magnetohydrodynamics is required to appropriately describe the far from equilibrium fluid known as the quark gluon plasma.

While it is possible to include magnetic fields, counted as $O(1)$ quantities in the derivative expansion~\cite{Grozdanov:2016tdf,Hernandez:2017mch}, it is unclear how large such fields can be before the description of the collective excitations, generated by hydrodyanmics, breaks down. Here it is important to again emphasize that strong refers to the derivative counting, strong fields are counted as $B\sim O(1)$, while the hydrodynamic description is expected to hold for $B/T^2 \ll 1$.  This distinction is especially relevant in the case of the CME in heavy ion collisions. As noted in~\cite{Fukushima:2008xe} for typical values~\footnote{These estimates are obtained in a weak field limit $B/T^2 \ll 1$ where the chiral chemical potential was found to scale as $\mu\sim n_5^{1/3}$ with a reasonable choice of magnetic field strength $eB\approx 10^4 \unit{MeV}$ just after the collision.} of the magnetic field strength, and chiral chemical potential $\mu_5$ (as deduced by typical sphaleron sizes in QCD) it may be reasonable to expect $\mu_5/T\sim 1/10 $ to $1$ with the ratio of the magnetic field to the temperature taking similar ranges $B/T^2\sim 1/10$ to $1$. This clearly seams to be at odds with the hydrodynamic regime of validity. It is therefore crucial to understand the limitations of the hydrodynamic expansion in this context.

Fortunately there has been much recent interest in understanding the boundaries of the effectiveness of hydrodynamics (see for instance~\cite{Withers:2018srf,Grozdanov:2019kge,Grozdanov:2019uhi,Abbasi:2020ykq,Jansen:2020hfd,Heller:2020hnq,Cartwright:2021qpp,Jeong:2021zsv}). The authors of~\cite{Grozdanov:2019uhi} provided a particularly useful characterization of the breakdown of hydrodynamic description. Their work focused on the collective excitations about an equilibrium state, obtained as linearized fluctuations of energy and momentum, referred to as hydrodynamic modes. Each of these modes obeys a dispersion relation, a relation between the frequency $\omega$ and the momentum $\mathbf{q}$, which takes the form
\begin{equation}\label{eq:general_dispersion}
    \omega(\mathbf{q})=\sum_{j=0}^\infty a_n \mathbf{q}^{j/m} \,,
\end{equation}
where $\mathbf{q}$ is the wave vector, the coefficients of the series take values in the complex numbers i.e. $a_n\in\mathbb{C}$, and the exponent $m\in\mathbb{N}$. While here the series was written to include an infinite number of contributions, the series will truncate when considering a hydrodynamic expansion including finitely many derivative contributions. Naturally, when handed a series expansion one can begin naive tests of convergence of the series. Typically, this requires a study of the asymptotic behavior of the coefficients $a_n$, which, while in principle can be computed, the expressions to obtain each $a_n$ can become analytically or numerically intractable. With this in mind the authors of~\cite{Grozdanov:2019uhi}, borrowing techniques from the theory of plane analytic curves demonstrated that the information about the convergence radius the hydrodynamic dispersion relations can be obtained from the 1-fold, or critical points, of the functions which implicitly define the dispersion relations.

In this work we seek to provide some insight into the behavior of the radius of convergence of dispersion relations of the form shown in Eq. (\ref{eq:general_dispersion}) when subjected to strong fields. We will be interested in the framework of strong field external hydrodynamics~\footnote{For a related study on the breakdown of magnetohydrodynamics in 2+1-dimensions by studying QNM mode collisions see~\cite{Jeong:2021zsv}.}, in particular we will be interested in this framework when the transport coefficients which encode the microscopic degrees of freedom take the values appropriate for $\mathcal{N}=4$ supersymmetric Yang-Mills (SYM) theory minimally coupled to $U(1)$ gauge theory~\footnote{This theory has in the past been demonstrated to capture essential magnetic aspects of real QGP physics~\cite{Endrodi:2018ikq}. }. In section~\ref{sec:holoModel} we introduce the holographic model beginning with the construction of anisotropic, asymptotically AdS, black brane solutions of Einstein-Maxwell theory, the gravitational dual to $\mathcal{N}=4$ SYM theory minimally coupled to $U(1)$ gauge theory. Following this we detail the construction of the perturbations of these black brane solutions, keeping in mind where hydrodynamic behavior occurs in the dual theory, and discuss the method used to compute the critical points associated with the hydrodynamic modes. In section~\ref{sec:results} we will display the results of this calculation. Finally, we close with a brief discussion and suggestions for further research.

%%%%%%%%%%%%%%%%%%%%%%%%
\section{Holographic setup}
~\label{sec:holoModel}
In this section we will discuss the holographic framework used to provide an example of the behavior of the radius of convergence of the hydrodyanmic expansion subjected to strong fields. The section is broken into three pieces, the first describes the asymptotically AdS geometry and the construction of black brane solutions with non-trivial magnetic charges which prepares a thermal state $\mathcal{N}=4$ SYM subjected to strong external magnetic fields. The second piece describes the fluctuation equations used to probe this geometry from which we can extract the quasinormal frequencies of the black brane geometry dual to the poles of retard Greens functions of energy momentum tensor and axial current in the dual field theory. The final piece concerns the techniques need to compute the critical points which can loosely considered to be a quasinormal mode problem with an additional boundary condition.

Since the topic of both the background geometry (see for instance~\cite{DHoker:2009ixq,DHoker:2009mmn,DHoker:2010zpp,Cartwright:2021hpv}) and the fluctuations~\footnote{QNM's in a related holographic model built to mimic the abelian part of the QCD anomaly from a bulk with a local $U(1)_A\times U(1)_V$ gauge symmetry were considered first in~\cite{Ammon:2016fru} and in particular include zero momentum with vector magnetic field.} (see for instance~\cite{Kovtun:2005ev,Janiszewski:2015ura,Ammon:2017ded,Ammon:2020rvg}) have been discussed in detail in previous works, we will attempt to keep this section to a minimum, directing the interested reader to the relevant resources for more information.

\subsection{Background geometry}
The gravitational theory that is dual to $\mathcal{N}=4$ SYM theory at large $N$ and large t' Hooft coupling $\lambda$, minimally coupled to global electromagnetic fields associated with a $U(1)$ subgroup of the R-symmetry, is Einstein-Maxwell-Chern-Simons theory~\cite{DHoker:2009ixq,DHoker:2009mmn,DHoker:2010zpp}. In particular one must consider asymptotically $AdS_5$ solutions of the Einstein-Maxwell-Chern-Simons (EMCS) theory as first described in~\cite{DHoker:2009ixq}. The action is given as the following
\begin{align}
    S&=\frac{1}{16\pi G_5}\left(\int{\exd^5x\sqrt{-g}\left(R-2\Lambda-\frac{L^2}{4}F^{\mu\nu}F_{\mu\nu}\right)}\right. \nonumber \\
    & \left.-\frac{\bar{\gamma}}{6}\int A\wedge F\wedge F\right) +S_{ct}\, ,
\end{align}
where $G_5$ is the five-dimensional Newton constant, $\Lambda=-6/L^2$ is the cosmological constant, $\bar{\gamma}$ is the Chern-Simons coupling and $L$ is the $AdS$ radius. From here onward we will set the $AdS$ radius to $L=1$ and take the Chern-Simons coupling~\footnote{In a companion paper we will address the case of $\bar{\gamma}\neq 0$.} $\bar{\gamma}=0$. The additional term in the action, $S_{ct}$, contains counterterms, required to 1) make the gravitational variational problem well posed and 2) kill terms which diverge near the AdS boundary~\cite{Taylor:2000xw}
\begin{align}
    S_{ct}&=\frac{1}{8\pi G_5}\int\exd^4 x\sqrt{\gamma}\left( K -\frac{1}{2L}\left(-6  -\frac{L^2}{2}R(\gamma)  \right)\right)\nonumber \\
    &+\frac{L^3}{64\pi G_5}\log(\epsilon)\int\exd^4 x\sqrt{\gamma_0}F_0^2\, ,\label{eq:counter_term_action}
\end{align}
In the counterterm action $K$ is the trace of the extrinsic curvature, and $\gamma$ is the induced metric on a constant $z=\epsilon$ hypersurface (with a small cut-off energy value $\epsilon$), $\gamma_0$ is the metric of the dual field theory and $F_0$ is the external field strength of the gauge field $A$ in the dual theory.

Varying the action with respect to the metric and the gauge field lead to the equations of motion
\begin{align}
    R_{\mu\nu}-\frac{1}{2}(R-2\Lambda)g_{\mu\nu}&=\frac{1}{2}\left(F_{\mu\alpha}\tensor{F}{_{\nu}^{\alpha}}-\frac{1}{4}g_{\mu\nu}F_{\alpha\beta}F^{\alpha\beta}\right)\, , \label{eq:einstein}\\
    \nabla_{\mu}F^{\mu\nu}&=0\, .
\end{align}
With our goal to study quasinormal mode like problems it is advantageous to make use of the infalling Eddington-Finkelstein like metric ansatz introduced in~\cite{Ammon:2017ded},
\begin{align}\label{eq:metric_ansatz}
    \exd s^2&=\frac{1}{z^2}\left(-2\exd z\exd \nu-U(z)\exd \nu^2+v(z)^2\left(\exd x_1^2+\exd x_2^2\right) \right.\nonumber\\
    &\left.+w(z)^2\exd x_3^2\right),
\end{align}
with coordinates $(\nu,x_1,x_2,x_3,z)$ where the $AdS$-boundary located at $z=0$. The gauge field ansatz is given by
\begin{equation}\label{eq:Gauge_Ansatz}
    A=\frac{B}{2}(-x_2\exd x_1+x_1\exd x_2) \\
\end{equation}
Insisting that there exist a timelike Killing vector which acts as the null generator of a surface $S=z-z_h$ in the bulk implies that there exists an event horizon whenever $U(z_h)=0$. Hence the ansatz chosen, along with these boundary conditions, leads to anisotropic black brane solutions. Furthermore, it is necessary to impose that near $z=0$ the metric takes the form
\begin{equation}
ds^2=\frac{1}{z^2}\left(\eta_{ij}\exd x^{i}\exd x^j\right)\, ,
\end{equation}
for $\eta=\text{diag}(-1,1,1,1)$, the Minkowski metric, to ensure the solution is asymptotically AdS. With this asymptotic behavior, the equations of motion can be solved order by order in an expansion in the AdS radial coordinates $z$. Up to $O(z^4)$ these solutions are given by
\begin{subequations}\label{eq:near_bndy_expansion}
\begin{align}
    U(z)&=1+ u_4 z^4 +\frac{B^2}{6}\log(z)z^4+\cdots \, ,\\
    v(z)&=1+ v_4 z^4 -\frac{B^2}{24}\log(z)z^4+\cdots \, , \\
    w(z)&=1+ w_4 z^4 +\frac{B^2}{12}\log(z)z^4+\cdots \, ,
\end{align}
\end{subequations}
 as have been detailed in past works (see for instance~\cite{Janiszewski:2015ura,Ammon:2016szz,Ammon:2017ded,Ammon:2020rvg,Cartwright:2021hpv}).
A standard holographic relation can be used to construct~\footnote{The coefficients given in eq.\ (\ref{eq:holo_Energy_Momentum}) can be extracted by transforming to the Fefferman-Graham coordinate system. Expanding the metric near the $AdS$ boundary gives~\cite{deHaro:2000vlm},
\begin{align}
    \exd s^2&=\frac{\exd\rho^2}{4\rho^2}+\frac{1}{\rho}g_{ij}(x,\rho)\exd x^i\exd x^j\, , \\
    g(x,\rho)&=g_{(0)}+\rho^2( g_{(4)}+h_{(4)}\log(\rho))+\cdots \, .
\end{align}}
the one-point functions of the dual energy-momentum tensor and global $U(1)$-current~\cite{Fuini:2015hba,DHoker:2009ixq,Ammon:2016szz}
\begin{align}
    \braket{T_{ij}}&=\frac{2}{\kappa^2}\left(g_{(4)ij}-g_{(0)ij}\tr g_{(4)}-(\log(\Lambda) +\mathcal{C})h_{(4)ij}\right) \, ,\label{eq:holo_Energy_Momentum} \\
     \braket{J^{\mu}}&=\frac{2}{\kappa^2}\left(\lim_{z\rightarrow 0}\frac{1}{z^3}\eta^{\mu\nu}\partial_zA_{\nu}\right)\, \label{eq:Consistent_Current}
\end{align}
where $\mathcal{C}$ is an arbitrary scheme-dependent constant and $\kappa^2=8\pi G_5$. In what follows we will take $G_5=1/16\pi$. Indeed the divergence of this current (i.e. the corresponding hydrodynamic equation of motion) is conserved
\begin{equation}
    \partial_\mu \braket{J^{\mu}}= 0
\end{equation}
when the Chern-Simons coupling is zero and is otherwise given as expected by the chiral anomaly~\cite{Bell:1969ts,Adler:1968av,Ammon:2016szz,Ammon:2020rvg}. Since the explicit values of the one-point function of the energy-momentum tensor~\footnote{The external gauge field explicitly violates conformal invariance
\begin{equation}
    \braket{\tensor{T}{^{a}_{a}}} = -\frac{1}{4}F_{ab}F^{ab} =-\frac{B^2}{2} \, .
\end{equation}
The violation is on the level of the state, not the theory.} will be useful to characterize the solutions we obtain we express the formulas in Eq. (\ref{eq:holo_Energy_Momentum}) and Eq. (\ref{eq:Consistent_Current}) in terms of the coefficients of the near boundary expansion of the metric given in Eq. (\ref{eq:near_bndy_expansion}) as
\begin{align}
    \braket{T^{tt}}&=-3u_4 -\frac{B^2}{2}\log(\Lambda_B),\\
    \braket{T^{ii}}&=-\frac{B^2}{4}-u_4-4w_4+\frac{B^2}{2}\log(\Lambda_B), \\
    \braket{T^{x_3x_3}}&=8w_4-u_4 -\frac{B^2}{2}\log(\Lambda_B),\\
      \braket{J^{\mu}}&=0
\end{align}
where $i=x_1,x_2$. The quantity $\Lambda_B$ is an energy scale, and its introduction is unavoidable in this system. It may be thought off as stemming from an arbitrary separation or partitioning of the full energy-momentum tensor of the system into an electromagnetic and SYM contribution. In this work we choose the energy scale $\Lambda_B=B^{1/2}$ (much more about the renormalization point dependence and choice of renormalization scale is discussed in~\cite{Fuini:2015hba,Grozdanov:2017kyl}).

Inserting the ansatz given in Eq. (\ref{eq:metric_ansatz}) and Eq. (\ref{eq:Gauge_Ansatz}) into the Einstein-Maxwell equations of motion leads to a coupled, nonlinear, set of five differential equations for three variables. After a small set of manipulations this can be reduced to a set of three second order differential equations and a single first order constraint. As discussed in~\cite{Cartwright:2021hpv} there exists only a small number of analytic solutions to these equations of motion and as a result we must resort to numerical methods to construct solutions. In this work the same techniques described in the appendix of~\cite{Cartwright:2021hpv} will be used to obtain numerical solutions to the equations of motion. As such we will not repeat these steps here. In brief, the unknown metric functions are discretized in terms of a $N$-th order truncated Chebysehv representation and are hence given by their value at $N$ Gauss-Lobatto grid points. Then, the non-linear system of equations can be solved by using a Newton-Raphson root finding technique to iteratively improve an initial guess of the metric functions. The algorithm continues until a suitable measure of convergence has been met. Labeling the field equations by $E^i_j$ with $i=1,\cdots 3$ indexing over the field equation, and $j=1,\cdots N$ indexing over the grid points, $E^i_j$ represents the residual value of the $i$-th field equation at the $j$-th grid point. The algorithm continues until
\begin{equation}
    \varsigma = \frac{1}{3N}\sum_{i=1}^3\sum_{j=1}^N |E^i_j| \leq  5\times 10^{-12}\, .
\end{equation}
In addition, we use the residual of the left over constraint equation as an additional monitor of the goodness of our solution, whose largest value over all grid points, routinely takes values on the order $O(e^{-20})$. The interested reader can find a detailed account of this method is given in~\cite{Cartwright:2021hpv}.

\begin{figure}[htbp]
    \centering
    \includegraphics[width=0.47\textwidth]{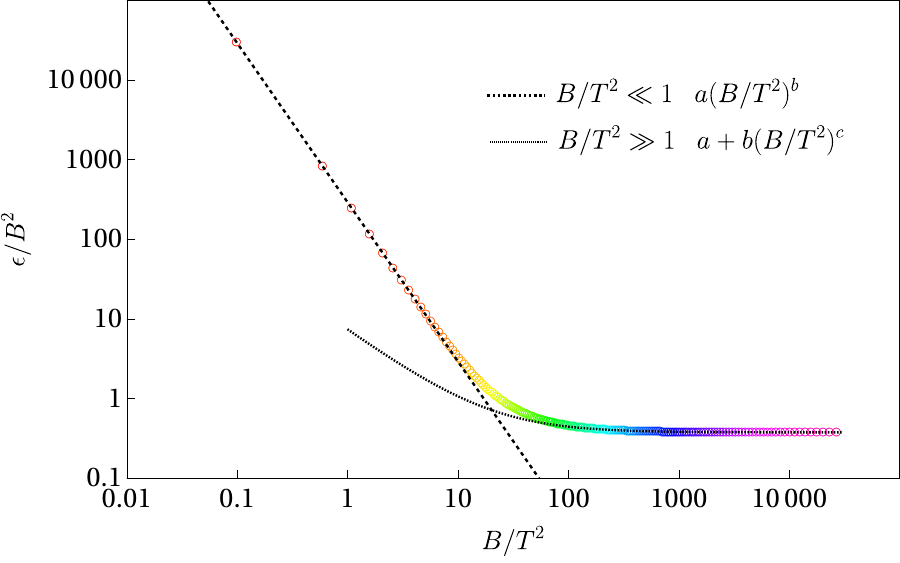}
    \caption{\textit{Magnetic Black Branes: }One parameter family of asymptotically AdS$_5$ solutions of Einstein-Maxwell theory. The lines corresponds to fits done at high and low temperature and the value of the fit parameters are displayed in table~\ref{tab:fitPara}.
    \label{fig:1pfamily}}
\end{figure}
The solutions constructed to the system of equations are governed by one dimensionless combination of the temperature ($T$) and the magnetic field ($B$). In this work we choose to use
\begin{equation}
    \bar{B}=\frac{B}{T^2}\, .
\end{equation}
To construct solutions at fixed values of these parameters requires an additional Newton-Raphson layer to the solver. One repeats the steps detailed above until solutions at a particular value of the dimensionless magnetic field $\bar{B}$ are obtained.

Having discussed ansatz and the methods for find solutions it is worth while to review the background solutions in a little bit more detail before closing the section. A characteristic curve describing the solutions used in this work is displayed in Fig.~\ref{fig:1pfamily} where a suitable measure of the energy density (scaled by the magnetic field) is plotted for various values of the dimension magnetic field (see for instance~\cite{Fuini:2015hba} for a further discussion). The transition between the low and high temperature behavior (large magnetic field and small magnetic field respectively) is somewhat dramatic. At zero magnetic there is clearly a planar Schwarzschild black brane is a solution to the equations of motion. What may not be so obvious is that there exists another analytic solution~\cite{DHoker:2009mmn} to this system~\footnote{There is some subtlety in the holographic interpretation associated with the coordinates. This is discussed in more detail in~\cite{DHoker:2009mmn}. } of equations, the product of a Ba\~{n}ados-Teitelboim-Zanelli (BTZ) black hole and a two dimensional manifold, BTZ$_{2+1}\times \mathbb{R}^2$, which in our conventions takes the form
\begin{equation}\label{eq:BTZ_IR_Metric}
   \exd s^2= -\frac{\exd \nu^2 \left(3-\frac{3 z^2}{z_h^2}\right)}{z^2}-\frac{2 \exd \nu \exd z}{z^2}+\frac{ B}{2 \sqrt{3}}\left(\exd x_1^2+\exd x_2^2\right)+3\frac{\exd x_3^2}{z^2} \, .
\end{equation}
where it should be noted that the effective AdS radius of the three dimensional theory is $l=L/\sqrt{3}$. Hence the solutions that we seek are those which interpolate between BTZ$_{2+1}\times \mathbb{R}^2$ at large $z$ and AdS$_{4+1}$ at small $z$ (the AdS boundary). One can interpret this as a renormalization group flow between a $3+1$ conformal field theory (CFT) in the UV and a $1+1$ CFT in the IR driven by the magnetic field. The UV theory in this case is $\mathcal{N}=4$ SYM in an external magnetic field and it has been shown that in the free field limit that this theory behaves as a $1+1$ CFT at low temperatures~\cite{DHoker:2009mmn}. Furthermore, from Fig.~\ref{fig:1pfamily} one sees that there is indeed a smooth, one parameter family of solutions interpolating between these two limits.

\begin{table}[htbp] \caption{Boltzmann relation between energy density and temperature in the magnetic black brane geometry.    \label{tab:fitPara}}
    \centering
    \begin{ruledtabular}
    \begin{tabular}{l | l |l|l| l }
    Regime & Fit function $\epsilon/B^2$ & $a$ & $b$ & $c$ \\
    \hline
    $B/T^2\ll 1 $ & $ a (B/T^2)^b$ & 292.206 & -2.000 & N/A \\
   $B/T^2\gg 1 $ &  $ a +b (B/T^2)^c$  &  0.374 & 7.027 & -1.007
    \end{tabular}
    \end{ruledtabular}
\end{table}

The change in the dimensionality of the field theory description can be seen clearly by investigating the thermodynamic data more carefully. Here we choose to demonstrate by fitting the relation between the energy density and temperature. For $\bar{B}\ll 1$ a fit $\epsilon/B^2= a (B/T^2)^b$ is appropriate, while for $\bar{B}\gg 1$ we use $\epsilon/B^2= a+b(B/T^2)^c$. The resulting parameters of the fit are described in table~\ref{tab:fitPara}. For $\bar{B}\ll 1$ one expects a Schwarzschild geometry, in our units, given by $\epsilon/B^2=3(\pi T)^4$. Indeed referring to the table the relative difference $2|a-(3\pi)^4|/(a+(3 \pi)^4)=7.135\times 10^{-5}$. While $b$ is the expected power to three decimals. Likewise for $\bar{B}\gg 1$ we find the expected power law behavior BTZ$_{2+1}$ discussed in~\cite{Fuini:2015hba}. Additionally, the coefficients $a$ and $b$ agree with theirs to one decimal. The likely reason for this disagreement is two fold, first, they choose a fit function for which the coefficient $c$ is fixed as $c=-1$. Here we allow the fitting routine to choose the best value for this parameter. Second is the range fitted, in their work they consider energy density and temperature relation $(\pi T)^4/B^2$ in the range $(10^{-3},10^2)$. While the range in this work corresponds to $(10^{-7},10^5)$ hence the fits we construct use data three orders of magnitude closer to the limit $\bar{B}\rightarrow\infty$. Hence what is displayed in this work should be considered more accurate then~\cite{Fuini:2015hba}.

%%%%%%%%%%%%%%%%%%%%%%%
\subsection{Fluctuations}
The poles of the retarded Greens functions can be captured in the resonant response of the anisotropic black brane geometries discussed in the previous section as originally detailed in~\cite{Birmingham:2001pj,Son:2002sd,Kovtun:2005ev}. These are referred to as quasinormal modes, and the calculation of quasinormal modes of the black brane geometries considered in this work have been discussed in detail in~\cite{Janiszewski:2015ura, Ammon:2017ded,Ammon:2020rvg}. The fluctuations are solutions to the linearized Einstein-Maxwell system and are parameterized as,
\begin{align}
g_{\mu\nu}&=g^{(0)}_{\mu\nu}+\epsilon g^{(1)}_{\mu\nu}+O(\epsilon^2)\, , \nonumber \\ A_\mu&=A^{(0)}_{\mu}+\epsilon A^{(1)}_{\mu}+O(\epsilon^2)\, ,
\end{align}
where $g^{(0)}$ and $A^{(0)}$ are the solutions obtained for the metric and gauge field discussed in the previous section. Then the $O(\epsilon)$ field equations take the form,
\begin{widetext}
\begin{align}
  0&=\frac{1}{2}\nabla_\mu \nabla_\nu g^{(1)}-\frac{1}{2}\nabla^\lambda \nabla_\lambda  g^{(1)}_{\mu\nu}+\nabla^\lambda \nabla_{(\mu}g^{(1)}_{\nu)\lambda}- \frac{2\Lambda}{D-2} g^{(1)}_{\mu\nu}-s(A^{(0)},A^{(1)})\,,  \label{eq:EH_Fluc}\\
 0&= \partial_{\mu}\left[ \sqrt{-g^{(0)}}\left(\frac{1}{2}\tr(g^{(0) \alpha \beta}g^{(1)}_{\beta\nu})F^{(0)\mu\nu}+F^{(1)\mu\nu}+\left(g^{(0)\mu\alpha}g^{(1)\nu\beta}+g^{(1)\mu\alpha}g^{(0)\nu\beta} \right)F^{(0)}_{\alpha\beta}\right)  \right]  \label{eq:maxwell_cs_Fluc}
\end{align}
\end{widetext}
where indices are raised and lowered by the order zero metric and the source term $s$ is given by,
\begin{widetext}
\begin{align}
    s(A^{(0)},A^{(1)})&=\frac{1}{2}\left(g^{(1)\lambda\beta} F^{(0)}_{\mu\lambda}F^{(0)}_{\nu\beta} + 2F^{(0)}_{(\mu\lambda}\tensor{F}{ ^{(1)}_{\nu)}^\lambda} \nonumber  \right. \\
   & \left. +\frac{1}{2(2-d)} \left( 2g^{(0)}_{\mu\nu}F^{(0)}_{\lambda\eta}F^{(1)\lambda\eta}   +g^{(1)}_{\mu\nu}F^{(0)}_{\lambda\eta}F^{(1)\lambda\eta}+2 g_{\mu\nu}^{(0)} F^{(0)}_{\lambda\beta}\tensor{F}{^{(0)}^\lambda_\eta}g^{(1)\beta\eta}\right) \right)
\end{align}
\end{widetext}
In constructing the fluctuation equations it is advisable to use the form given by Eq. (\ref{eq:EH_Fluc}) and Eq. (\ref{eq:maxwell_cs_Fluc}) to avoid lengthy expressions which might otherwise be obtained by direct brute force substitution of the expansion ansatz into the equations of motion.

The solutions to the background equations of motion are stationary spacetimes, hence it is highly convenient to work in a Fourier representation of the fluctuating fields $g^{(1)}$ and $A^{(1)}$,
\begin{align}
    g^{(1)}_{\mu\nu}(z,x^i)&=\int \exd^4 k e^{-i k_i x^i}g^{(1)}_{\mu\nu}(z,k^i)\, , \\
    A^{(1)}_{\mu}(z,x^i)&=\int \exd^4 k e^{-i k_i x^i}A^{(1)}_{\mu}(z,k^i)
\end{align}
where we explicitly make the gauge choice $g^{(1)}_{z\mu}=0$ and $A^{(1)}_z=0$. Furthermore we will focus on the case of $k||B||x_3$, leaving a little group of rotations $SO(2)$ under which we can group the fluctuations into sectors. These sectors are displayed in table~\ref{tab:spin_sectors}.

\begin{table}[h]
 \caption{The decomposition of the fluctuations $g^{(1)}$ and $A^{(1)}$ in terms of the little group $SO(2)$, along with each sectors transformation under the remaining discrete symmetries is displayed. Further information about this decomposition can be found in~\cite{Ammon:2017ded}.
    \label{tab:spin_sectors}}
  \begin{ruledtabular}
    \begin{tabular}{llll}
       \text{Spin}  & \text{Field content}  &  $\begin{cases}
B\rightarrow -B \\
\gamma\rightarrow -\gamma
\end{cases}$  &   $\begin{cases}
B\rightarrow -B \\
k\rightarrow -k
\end{cases}$  \\
       \hline
      $2^{+}$   &  $g^{(1)}_{x_1x_2}\, , \, g^{(1)}_{x_1x_1}-g^{(1)}_{x_2x_2}$ & $2^{+}\rightarrow 2^{+}$  & $2^{+}\rightarrow 2^{+}$  \\
      \hline
       $1^{\pm}$   & $g^{(1)}_{\nu x_{\pm}}\, , \, g^{(1)}_{x_3x_{\pm}}\, , \, A^{(1)}_{x_{\pm}} $ & $1^{\pm}\rightarrow 1^{\mp}$ &$1^{\pm}\rightarrow 1^{\mp}$\\
       \hline
       \multirow{2}{*}{ $0^{+}$}  &  $g^{(1)}_{\nu\nu}\, , \, g^{(1)}_{x_1x_1}+g^{(1)}_{x_2x_2}\, , \, g^{(1)}_{x_3x_3}\,$  & \multirow{2}{*}{  $0^{+}\rightarrow 0^{+}$}& \multirow{2}{*}{ $0^{+}\rightarrow 0^{+}$ }\\
       & $g^{(1)}_{\nu x_3}\, , \, A^{(1)}_{\nu}\, , \, A^{(1)}_{x_3}$ & &
    \end{tabular}
    \end{ruledtabular}
\end{table}
Inserting the Fourier ansatz into the field equations (Eq. (\ref{eq:EH_Fluc}) and Eq. (\ref{eq:maxwell_cs_Fluc})) one naturally finds that the differential equation for $g^{(1)}_{\mu\nu}(z,k^i)$ and $A_{\mu}^{(1)}(z,k^i)$ may be represented as a generalized eigenvalue problem
\begin{equation}\label{eq:Gen_Eigen_Problem}
M(\w,\q^2)\Phi = (M_0(\q^2)+\w M_1(\q^2)+\w^2 M_2(\q^2)\cdots)\Phi=0\, .
\end{equation}
Here $M$ is a differential operator acting on the field content $\Phi$. The operator $M$ can be expanded as a power series, with coefficients $M_j$, in the frequency $\w$ and each coefficient may be expanded in the operator $d/dz$ truncating at second order
\begin{equation}
    M_j=M_j^{(0)} I+M_j^{(1)}\frac{\exd }{\exd z}+M_j^{(2)}\frac{\exd^2 }{\exd z^2}\, .
\end{equation}
Furthermore as expected, the operator $M$ and the field content $\Phi$ can be grouped as in table~\ref{tab:spin_sectors} leading to the generalized eigenvalue problem splitting into 4 decoupled sectors. In each sector the differential operator associated with that sector encodes the quasinormal mode spectrum.

The simplest case is if the operator is linear in $\w$, then one can directly solve the linear generalized eigenvalue problem (albeit numerically). If the power series expansion of $M(\w,\q^2)$ terminates at powers of $\w^2$ or higher then one again can solve directly at the expense of increasing the field content, e.\ g.\ for $\w^2$ we would have
\begin{align}
  \tilde{A} \tilde{\Phi}&=-\w \tilde{B}\tilde{\Phi}\, , \nonumber\\
  \tilde{A}&=\begin{pmatrix}
    M_0 & M_1 \\
    0 & I
    \end{pmatrix},\quad \tilde{B}=\begin{pmatrix}
    0 & M_2 \\
    -I & 0
    \end{pmatrix}
\end{align}
where $\tilde{\Phi}$ is a vector containing all fields and their associated auxiliary fields $\tilde{\Phi}=(\Phi,\Phi_1)$. Expanding this system of equations one finds that one of the equations simply impose that $\Phi_1=\w\Phi$. Back substitution into the remaining equation leads directly to Eq.~\ref{eq:Gen_Eigen_Problem} with $M=M_0+\w M_1+\w^2 M_2$. For the case of the sound sector in the magnetic black brane geometry of interest, the operator $M$ truncates at $\w^2$.

%%%%%%%%%%%%%%%%%%%%%%%%%%%%
\subsection{Determination of the spectral curve}
\label{sec:Det_Method}
Our goal in this section will be to study the set of points determined as follows
\begin{equation}\label{eq:criticalMomentum}
  \left\{ (\q,\w) \left| P(\q,\w)=0\, ,  \partial_\w P(\q,\w)=0 \right. \right\}  \, .
\end{equation}
where we will work with the dimensionless momentum $\q=q/(2\pi T)$ and frequency $\w=\omega/(2\pi T)$. These points are referred to as the critical points of the spectral curve. Linearized fluctuations around equilibrium in a fluid are referred to as hydrodynamic modes e.g. Eq. (\ref{eq:general_dispersion}) and the equations of motion obeyed by such fluctuations can obviously be cast in the form of linear set of equations. A necessary condition for the existence of a nontrivial solution to the linear system is the vanishing of the determinant which defines an implicit function $P(\q,\w)=0$, a curve in the space $\mathbb{C}^2$.
The key insight of~\cite{Grozdanov:2019uhi} was the use of the theory of plane curves to discuss the process of solving for the resolution of the curve. In particular, for non-singular points the analytic implicit function theorem provides an algorithm for constructing local solutions around $(\q_0,\w_0)$ to $P(\q,\w)$ (see for instance~\cite{walker:1950alg,wall_2004}). As discussed further in~\cite{Amano:2023bhg} the most interesting aspect of~\cite{Grozdanov:2019uhi} is that one can continue to construct local solutions around $(\q_0,\w_0)$ to $P(\q,\w)$ even for 1-fold or critical points. These are the points for which the implicit polynomial function $P$, of say order $n$, satisfies $P(\q,\w)= \partial_\w P(\q,\w)=0$ but $\partial^n P(\q,\w)\neq 0$. Around such points, under the assumption that $P$ is analytic at $(\q,\w)$, a local solution with $n$-branches is guaranteed to exist.

Despite the simplicity of the result provided in~\cite{Grozdanov:2019uhi} computing this in practice can be difficult. Asides from the obvious dependence on the specific theory, simply constructing the implicit function can itself be a daunting task. It is here we turn to holography to provide a location where the construction of the implicit function can be done explicitly. As discussed in the previous section an efficient method of computing the quasinormal mode spectrum is the decomposition of the fluctuation equation in a generalized eigenvalue problem. However, while this is useful to obtain the quasinormal mode spectrum in an efficient way there is an alternative method which is much more useful for the calculation of the spectral curve and hence the critical points of the spectrum. The determinant of the operator itself provides a representation of the hydrodynamic spectral curve, $P(\q,\w)=\det(M(\q,\w))$. By constructing a discrete representation of the operator $M$, we have direct access to a discrete representation of the spectral curve.

Following~\cite{Cartwright:2021qpp} the background geometry constructed in section~\ref{sec:holoModel}, representing a thermal state of $\mathcal{N}=4$ SYM subjected to external fields, is already represented by a discrete set of values at a collection of $N$ Gauss-Lobatto grid points as part of a truncated pseudo-spectral representation~\footnote{In practice the number of grid points used in the representation of the fields used to obtain solutions for the background geometry can be different then that used in the fluctuation equations. When this occurs one can create an interpolation following~\cite{boyd}. Consider two grids, for the first $x_i=\cos(\pi i/N)$ with $i=0,\cdots N$ and for the second $y_i=\cos(\pi i/M)$ with $i=0,\cdots M$. One can construct cardinal functions given by $C_j(x)=(2/Np_j)\sum_{m=0}^N(1/p_m)T_m(x_j)T_m(x)$ where $T_m$ are the Chebyshev polynomials of the first kind and $p_j=2$ if $j=0$ or $N$ and is $p_j=1$ otherwise. A function $f$ expanded in the cardinal basis can then be represented by $f(x)=\sum_{k=0}^{N}f(x_k) C_k(x)$ where $f(x_k)$ are the values of the function at each grid point. One can now obtain the function on the new grid $y_j$ by using the cardinal representation $f(y_j)=\sum_{k=0}^{N}f(x_k) C_k(y_j)$.}. One then also performs an expansion of the fields in the fluctuation equations in a truncated pseudo-spectral representation. For $n$ fluctuations and $N$ grid points this implies that the operator $M$ is a, dense,  $nN\times nN$ matrix. Following~\cite{Cartwright:2021qpp} the matrix $M$ can be simplified in a $LU$ decomposition leading to the spectral curve being given by
\begin{equation}\label{eq:lu}
P(\q,\w)=\det (M(\q,\w)) = \det (LU)=\det (U)
\end{equation}
since $\det (L)=1$. This is computationally viable since $U$ is an upper triangular matrix whose determinant is simply the product of the diagonal entries. Critical points of the curve, defined by Eq.~\ref{eq:criticalMomentum} may then be obtained by a two dimensional, damped, Newton-Raphson method
\begin{equation}
    \vec{\xi}_{k+1}=\vec{\xi}_{k}- \alpha J^{-1}\cdot \vec{X} \, ,
\end{equation}
where we have defined the following quantities
\begin{equation}
    \vec{\xi}=(\w,\q)\, , \quad \vec{X}=(P,\partial_\w P) \, ,
\end{equation}
with the Jacobian given by
\begin{equation}
   J=\begin{pmatrix}
        \partial_\w P & \partial_{\q}P \\
        \partial_\w^2P & \partial_\w\partial_{\q}P
    \end{pmatrix} \, .
\end{equation}
In order to facilitate the derivatives we make use of finite differences as in~\cite{Cartwright:2021qpp}
\begin{align}
\frac{\partial P }{\partial \w}&= \frac{P(\q,\w+\delta)-P(\q,\w)}{\delta}\, , \\
\frac{\partial P }{\partial\q}&= \frac{P(\q+\delta,\w)-P(\q,\w)}{\delta}\, ,\\
\frac{\partial^2 P }{\partial \w^2}&= \frac{P(\q,\w+2\delta)-2P(\q,\w+\delta)+P(\q,\w)}{\delta^2} \, ,\\
\frac{\partial^2 P }{\partial \w\partial\q}&= \frac{1}{\delta^2}\left(P(\q+
\delta,\w+\delta)-P(\q+\delta,\w) \right. \nonumber\\
& \left. -P(\q,\w+\delta)+P(\q,\w)\right)\, ,
\end{align}
where we take $\delta=10^{-15}$. The coefficient $\alpha$ is the damping factor, it's value is determined based on the residual and is taken to be \begin{equation}
    \alpha=\begin{cases}
        1/5 & \frac{1}{2}\vec{X}^\dagger\vec{X}> 1/5 \, ,\\
        1 & \frac{1}{2}\vec{X}^\dagger\vec{X}< 1/5 \, .
    \end{cases}
\end{equation}
The value chosen for $\alpha$ is arbitrary, and can be adjusted to assist convergence of the method. Each step is initiated with a guess for the value of $\vec{\xi}_k$ and continues until either the residual is less then $10^{-6}$ or the method stalls and the difference between each step is less then $10^{-25}$. The entries of the spectral derivative matrices become increasingly large as the number of grid points is increased. As a result the values obtained from Eq.~\ref{eq:lu} become enormous quantities and obtaining small numbers, on the order of $10^{-6}$ as required by the scheme requires a nearly miraculous cancellation of big numbers. This is why we work with a criterion based on when the method slows, and the correction to $\vec{\xi}$ becomes small. One can envision the surface spanned by $P$ in the complex momentum and frequency plane, and we hope to find places where the value that the function takes is significantly smaller then average value of the surface at a generic point. Hence, we require a regularization of the determinant in order to make meaningful statements of about the relative size of the value of $P$ evaluated at some location in the complex plane. A simple way to do this, and obtain some relative information about the magnitude of the spectral curve is to pick an arbitrary value of the frequency and momentum $(\q_{ref},\w_{ref})$ and evaluate the curve. Then one can compose $P(\q,\w)/P(\q_{ref},\w_{ref})$. Doing so one finds that, for instance at a critical point $(\q_c,\w_c)$ calculated for $B/T^2\approx 1/10$, the regulated curve takes a value $P(\q,\w)/P(\q_{ref},\w_{ref})\approx 10^{-17}$. In practice, we find that simply taking a the curve at $P(1,1)$ is enough to generate a reference value in which we can meaningfully compare the value of the curve at the critical point to the value of the curve in general.

With a method to construct the spectral curve, we can now obtain critical points. However, it is worth first understanding how to interpret the results, and in particular discuss a subtly of the analysis. As discussed in \cite{Grozdanov:2019kge,Grozdanov:2019uhi,Grozdanov:2021jfw}, when the order $n>1$ the equation $P(\q,\w)=0$ has multiple solutions, it is at these locations that a hydrodynamic mode ``collides'' with a gapped mode. If it is the case that this singular behavior of the curve is associated with a branch point, this is referred to as a ``level-crossing'' event. In these situations one can demonstrate, with an explicit example, using the Puiseux series (such as Eq.~\ref{eq:general_dispersion}) that the dispersion relation for the $i$-th mode behaves locally as $\w_i \sim (\q^2-\q_c)^{v}$ where $v$ is a fractional power. The fractional exponent $v=1/2$ is especially common for complex curves describing hydrodynamic dispersion relations. However, there also exists the possibility of locally analytic branches where $v$ is a positive integer, in the literature this is referred to as ``level-touching". The nearest critical point associated with a branch point singularity, say $\q_c$, limits the radius of convergence of the series $\w_i(\q^2)$ around the point $\q_0^2$ to the distance between the point of expansion and the critical point, $R=|\q_0^2-\q_c^2|$.

Since in practice we obtain the critical points numerically, it is important to understand how to differentiate numerically between ``level-crossing'' and ``level-touching''. One way of doing this is to use the method discussed in ~\cite{Grozdanov:2021jfw}, studying the monodromy of the curve. Practically we can study the trajectory of the frequency $\w(|x_i|^2 e^{i\phi})$ with $i=1,2$ for $|x_1|<|\q_c|<|x_2|$ and $\phi\in[0,2\pi]$. For values of the momentum below and above the critical momentum the trajectory of the frequencies associated with a ``Level-touching'' event while remain closed loops under phase rotation of the momentum. However, the frequencies associated with a ``Level-crossing'' event, while remaining closed loops under phase rotation of the momentum below the critical value, will merge into a single trajectory for values of the momentum above the critical momentum. Hence each branch of the frequency is mapped into itself under monodromy (an excellent image of this occurring in the simple example, used also in this work, is given in~\cite{Grozdanov:2021jfw}).

While we will display some images of this process occurring (see for instance Fig.~\ref{fig:monodromy_1} and Fig.~\ref{fig:monodromy_19}), it unfortunately becomes very difficult to obtain in this system for large values of the magnetic field. Increasing the magnitude of the magnetic field requires larger and larger grid sizes (due to the logarithms which appear in the near boundary expansion) to obtain even the first QNM reliably. See for example~\cite{Ammon:2017ded} where it is shown that even for the first three QNM of the helicity $0^+$ sector grids of size $N=100$ were used. This leads to a dense $800\times 800$ matrix of which one needs to obtain generalized eigenvalues (we find the same thing ocuring in our numerics, leading us to difficultly in finding the solutions to the generalized eigenvalue problem at large magnetic field strength). Hence it quickly becomes computationally infeasible to compute the generalized eigenvalue problem for $M$ discrete $\phi$ values in the interval $[0,2\pi]$ above and below the critical value at large magnetic field. Therefore, we need a means of determining whether the critical points we obtain are the result of level-crossing or level-touching. Fortunately, we appeal to the definition of the critical points to instruct us in this regard.

Recall that 1-fold points obey $P(\q,\w)=0$, and $\partial_\w P(\q,\w)=0$ but $\partial^n P(\q,\w)\neq 0$. In particular, for a solution to be a 1-fold point~\cite{walker:1950alg}
\begin{equation}\label{eq:diff}
    \partial_\q P(\q,\w)\neq 0 \, .
\end{equation}
One can use this criteria to differentiate level-crossing from level-touching.
\begin{definition}[Level-touching points]
    A level-touching point is a solution to $P(\q,\w)=0$, given by $(\q_0,\w_0)$ for which $\partial_\w P(\q_0,\w_0)=0$ and $\partial_\q P(\q_0,\w_0)=0$. In fact this is what is referred to as a \textit{singular point} of the curve.
\end{definition}
For this reason, level-touching really should not be considered as critical points of the curve, at least by the definition of 1-fold points given by Walker~\cite{walker:1950alg}. Likewise
\begin{definition}[Level-crossing points]
    A level-crossing point is a solution to $P(\q,\w)=0$, given by $(\q_0,\w_0)$ for which $\partial_\w P(\q,\w_0)=0$ and $\partial_\q P(\q_0,\w_0)\neq 0$.
\end{definition}
Hence, a critical point is identical to the notion of 1-fold point. Surprisingly, this simple means of differentiating level-crossing versus level-touching has not appeared in the literature (to the authors knowledge). That Eq.~\ref{eq:criticalMomentum}  provides a means of checking level-crossing versus level-touching can be seen readily in a simple example (originally given in~\cite{Grozdanov:2021jfw} and we repeat them here for illustrative purpose). The curves we compare are given by
\begin{subequations}
\begin{align}
    P_1(x,y)&= a^2-b^2+2bcx^2-c^2x^4-2ay+y^2\, ,\\
    P_2(x,y)&=a^2-b+cx^2-2ay+y^2  \, ,
\end{align}
\end{subequations}
where $x,y\in\mathbb{C}$. It can be easily checked that $(x_0,y_0)=(\pm \frac{\sqrt{b}}{\sqrt{c}},a)$ satisfy the requirements of Eq.~\ref{eq:criticalMomentum}. Furthermore, one can check immediately that,
\begin{subequations}
    \begin{align}
    \partial_xP_1(x,y)&= 4bcx-4c^2 x^3 \Longrightarrow \partial_xP_1(x_0,y_0)=0\, , \\
    \partial_xP_2(x,y)&= 2cx\Longrightarrow \partial_xP_2(x_0,y_0)=-2\sqrt{b c}\, .
    \end{align}
\end{subequations}
By our definitions it is clear that the location $(x_0,y_0)$ is an example of level-touching (for the curve $P_1$) and level-crossing (for the curve $P_2$). Furthermore it is simple to see this visually, by means of the monodromy of the curve.

A more physically relevant example of the application of these definitions is for the BTZ black hole~\cite{Banados:1992wn} (this case was also studied as an example in~\cite{Grozdanov:2019kge}). In particular the authors consider the case of the finite-temperature retarded two-point functions of operators with operator dimension $\Delta$ and spin $s=0$ in $2d$ CFT~\footnote{Notice, the energy-momentum tensor has $\Delta=s=2$ and hence is not of this type.}. The expression for these correlation functions is known analytically~\cite{Birmingham:2001pj,Son:2002sd}, hence one can extract the poles of the correlator in closed form and they are given by
\begin{equation}\label{eq:BTZ_disp}
    \w(\q)=\pm \q -i (2n+\Delta)\, , \quad n\in\mathbb{Z} \, .
\end{equation}
Extending the value of $\q \in\mathbb{C}$ these can be plotted for $\q=|\q| e^{i\phi}$ as the phase varies from $0$ to $2\pi$ to see that the plus/minus pair form circles in the complex frequency plane. And, as the authors argue it is clear from the plots that the minus branch will collide with the positive branch when $\w^-_{n}=\w^+_m$ for $m\neq n$ (they consider also the case with $n=0$ or $m=0$ separately) leading to a closed form for the critical points
\begin{align}\label{eq:BTZ_crit}
    \q_c&=i(m-n) \, ,\\
    \w_c&=-i(m+n+\Delta)\, .
\end{align}
What is not clear from their discussion is how they obtained this (clearly from the observation obtained from the plots one can set $\w_c=\w^-_{n}=\w^+_m$ and solve for $\q_c$ and substitute back for $\w_c$), and in particular if they used the technology they developed earlier. A simple way to obtain this is by realizing that the spectral curve associated with two modes colliding is given by the product of the implicit functions describing the two modes. The dispersion relation given in Eq.~\ref{eq:BTZ_disp} can be rewritten as
\begin{equation}
    P_n(\q,\w)=\w^2+2i\w(2n+\Delta)-(2n+\Delta)^2 -\q^2=0\, ,
\end{equation}
and hence the curve for a mode $m$ and $n$ interacting is given by
\begin{equation}
    P_{n,m}=P_n(\q,\w)P_m(\q,\w)=0\, .
\end{equation}
Applying the criterion to obtain critical points one must solve simultaneously
\begin{subequations}
\begin{align}
  P_{n,m}&=P_n(\q,\w)P_m(\q,\w)=0\, ,\label{eq:two_modes_critical_1} \\
     \partial_\w P_{n,m}&=\partial_\w P_n(\q,\w)P_m(\q,\w)+P_n(\q,\w)\partial_\w P_m(\q,\w)=0\, .\label{eq:two_modes_critical_2}
\end{align}
\end{subequations}
It is simplest to first solve the second of the two equations for $\w$. There are three solutions given by
\begin{subequations}
    \begin{align}
        \w_c^{\pm} &= -i (m+n+\Delta) \pm  \sqrt{-(m-n)^2+\q^2} \, ,\label{eq:btzPnm_1}\\
        \w_c &= -i (m+n+\Delta ) \, .\label{eq:btzPnm_2}
    \end{align}
\end{subequations}
Each of these may be propagated to the first equation to solve for $\q$. The first of these (Eq.~\ref{eq:btzPnm_1}) leads to $4 \q^2 (m-n)^2=0$. Hence either $n=m$ which contradicts our initial assumption, so we discard it, or $\q_c=\q=0$. In this case it says that zero momentum there are a collection of modes,
$\w^{\pm}=-i (\mp| m-n| +\Delta +m+n)$, aligned along the imaginary axis, which coincide and satisfy Eq.~\ref{eq:criticalMomentum}. In the other case (Eq.~\ref{eq:btzPnm_2}), one finds $\q_c=i(m-n)$, precisely as found in~\cite{Grozdanov:2019kge}.

With the spectral curve which reproduces these past results (and finds new modes) one can now check what the value of $\partial_q P_{n,m}(\q,\w)$ is and apply the definitions for level-touching and crossing. In both cases (Eq.~\ref{eq:btzPnm_1} and Eq.~\ref{eq:btzPnm_2}) one finds,
\begin{subequations}
    \begin{align}
        \partial_q P_{n,m}(\q_c,\w^{\pm}_c) =0\, ,\\
        \partial_q P_{n,m}(\q_c,\w_c)=0
    \end{align}
\end{subequations}
and hence we can say immediately these are not critical points, rather they are \textit{singular points}, and as a result correspond to \textit{level-touching}.

Clearly the spectral curve of the BTZ black hole has more then two modes interacting. That is, in general one finds the poles of the correalator can be expressed as
\begin{equation}
P_{BTZ}(\q,\w)=\prod_{i=0}^{\infty}P_i(\q,\w) = 0
\end{equation}
However, one can see quickly that singling out the $m$th and $n$th mode as
\begin{equation}
    P_{BTZ}(\q,\w)=P_m(\q,\w)P_n(\q,\w)\prod_{i\neq m, i\neq n}^{\infty}P_i(\q,\w) = 0
\end{equation}
leads to the critical point conditions
\begin{subequations}
\begin{align}  P_m(\q,\w)P_n(\q,\w)\prod_{i\neq m, i\neq n}^{\infty}P_i(\q,\w) &= 0\, ,\label{eq:full_curve_critical_1}\\
\left[\partial_\w P_m P_n +P_m\partial_\w P_n\right]\left(\prod_{i\neq m, i\neq n}^{\infty}P_i(\q,\w)\right)& \nonumber  \\
+P_m P_n \partial_w\left(\prod_{i\neq m, i\neq n}^{\infty}P_i(\q,\w)\right)&=0\, ,\label{eq:full_curve_critical_2}
\end{align}
\end{subequations}
one can see that provided that Eq.~\ref{eq:two_modes_critical_1} and Eq.~\ref{eq:two_modes_critical_2} are simultaneously satisfied, so to will Eq.~\ref{eq:full_curve_critical_1} and Eq.~\ref{eq:full_curve_critical_2} be simultaneously satisfied. In fact in this case $P_n(\q_c,\w_c)=P_m(\q_c,\w_c)=0$, i.e. each is separately zero. Furthermore, the additional condition that the point be a singular point $\partial_\q P_{BTZ}=0$ is identical in form to Eq.~\ref{eq:full_curve_critical_2} with $\partial_\w\rightarrow \partial_\q$ and since $\partial_{q}(P_m(\q,\w)P_n(\q,\w))=0$ and $P_m=P_n=0$ this is also satisfied in general when considering the full spectral curve.

Having demonstrated that the criterion makes sense, we finally state our method of monitoring the difference between level-crossing and level-touching numerically. When solving Eq.~\ref{eq:criticalMomentum} numerically, we are not able to demonstrate that each condition is zero. Rather, we are able to show that the value of the residue is small. In the numerical experiment we set a threshold of what, numerically, is zero. In this case, using 60-digit precision, the regulated residual is solved to a part in $10^{-17}$ (or smaller depending on the value of the magnetic field strength). Hence, numerically, this will serve as our threshold for zero. During the determination of the critical points we also check the value of $\partial_\q P(\q,\w)$. If it is above this threshold, we consider this value finite, if it is below the threshold, we consider it to be zero.

Finally, before we close this section it is important to understand in which sector we have the ability to check the radius of convergence of the hydrodynamic expansion. The discussion so far has centered on expansions of the curve around the point $(\q,\w)=(0,0)$. It is precisely these modes, for which $\lim_{\q\rightarrow 0}\w(\q)=0$, that are considered as hydrodynamic. Hence in what follows the distance from the origin in the complex plane to the nearest critical point associated with a branch point will set the radius of convergence. Table~\ref{tab:spin_sectors_Hydro} displays which sectors have hydrodynamic modes. It has been shown in~\cite{Ammon:2017ded} that the magnetic black branes we consider have gapped excitations in the $1^{\pm}$ and $2^+$ sector. Hence we will only consider the $0^+$ in this work. Further details on the behavior of hydrodynamic modes in the $0^+$ channel can be found in~\cite{Ammon:2017ded}.
\begin{table}[h]
 \caption{In the presence of the magnetic field only a single sector displays hydrodynamic poles.
    \label{tab:spin_sectors_Hydro}}
  \begin{ruledtabular}
    \begin{tabular}{ll}
       \text{Spin}  & Hydrodynamic modes\\
       \hline
      $2^{+}$  & \ding{55} \\
      \hline
       $1^{\pm}$  &\ding{55} \\
       \hline
        $0^{+}$ &\checkmark
    \end{tabular}
    \end{ruledtabular}
\end{table}

%%%%%%%%%%%%%%%%%%%%%%%%%%%%%
\section{Modes in the complex momentum space}\label{sec:results}
\begin{figure}
    \centering
    \includegraphics[width=0.47\textwidth ]{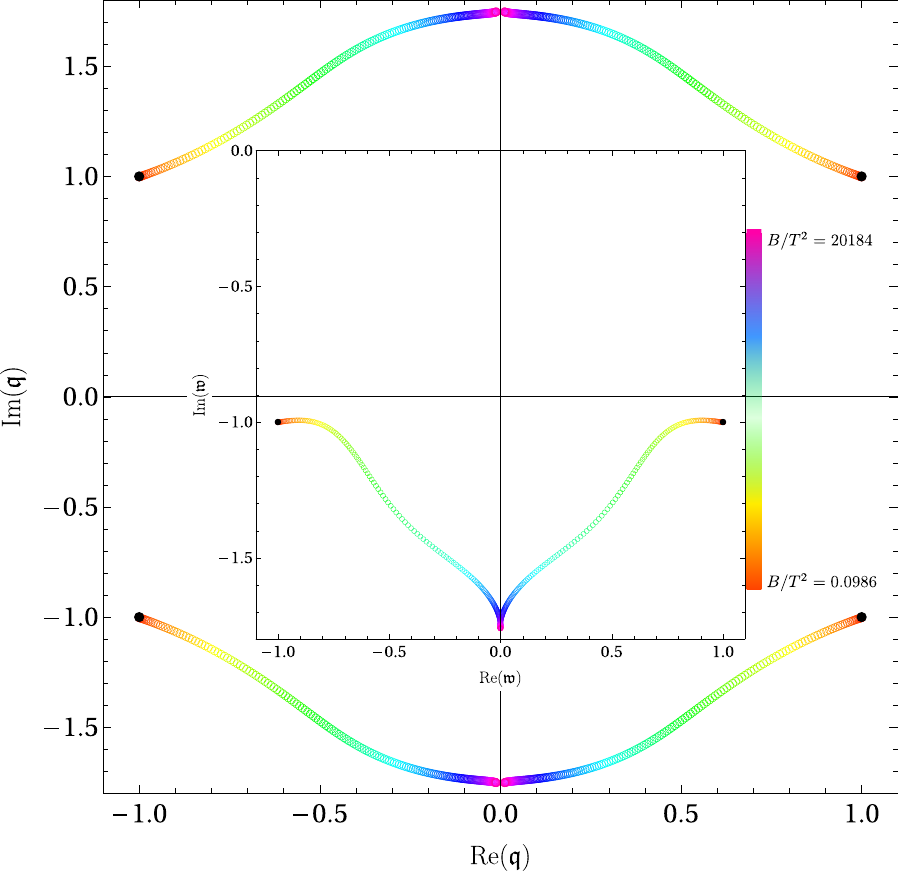}
    \caption{\textit{Complex Momentum:} The complex momentum plane is displayed with modes which satisfy Eq.~\ref{eq:criticalMomentum} in the linearized hydrodynamic expansion of $\mathcal{N}=4$ SYM minimally coupled to a global $U(1)$ gauge theory. The image displays the complex momentum plane, while the inset graphic displays the complex frequency plane. The colors of the plot display the strength of the magnetic field ($B/T^2)$. The black dots display the result at $B=0$.
    \label{fig:Complex_q_plane}}
\end{figure}

The result of the calculation described in section~\ref{sec:Det_Method} is displayed in Fig.~\ref{fig:Complex_q_plane}.
In the regime where $B/T^2\ll 1$, the hydrodynamic regime, the closet mode to the origin in the complex plane is a small perturbation away from the Schwarzschild result. This is of course expected, and is similar to the case of electrically charged fluids dual to Ressiner-Nordstr\"om black branes~\cite{Abbasi:2020ykq,Jansen:2020hfd}). In Fig.~\ref{fig:Complex_q_plane} the smallest value of the dimensionless magnetic field considered was $B/T^2\approx 1/10$, and is displayed as the black dots. Although $B/T^2\approx 1/10$ is not necessarily within the hydro regime, the relative difference ($2 \left| a-b/(a+b)\right|$) between the frequency and momentum for $B/T^2\approx 1/10$ and $B/T^2=0$ is on the order of $10^{-4}$. Already this is an interesting result suggesting that the radius of convergence of the dispersion relations remains relatively unchanged already as one approaches the boundary of the hydro regime.

\begin{figure}[htbp]
    \centering
    \includegraphics[width=0.47\textwidth ]{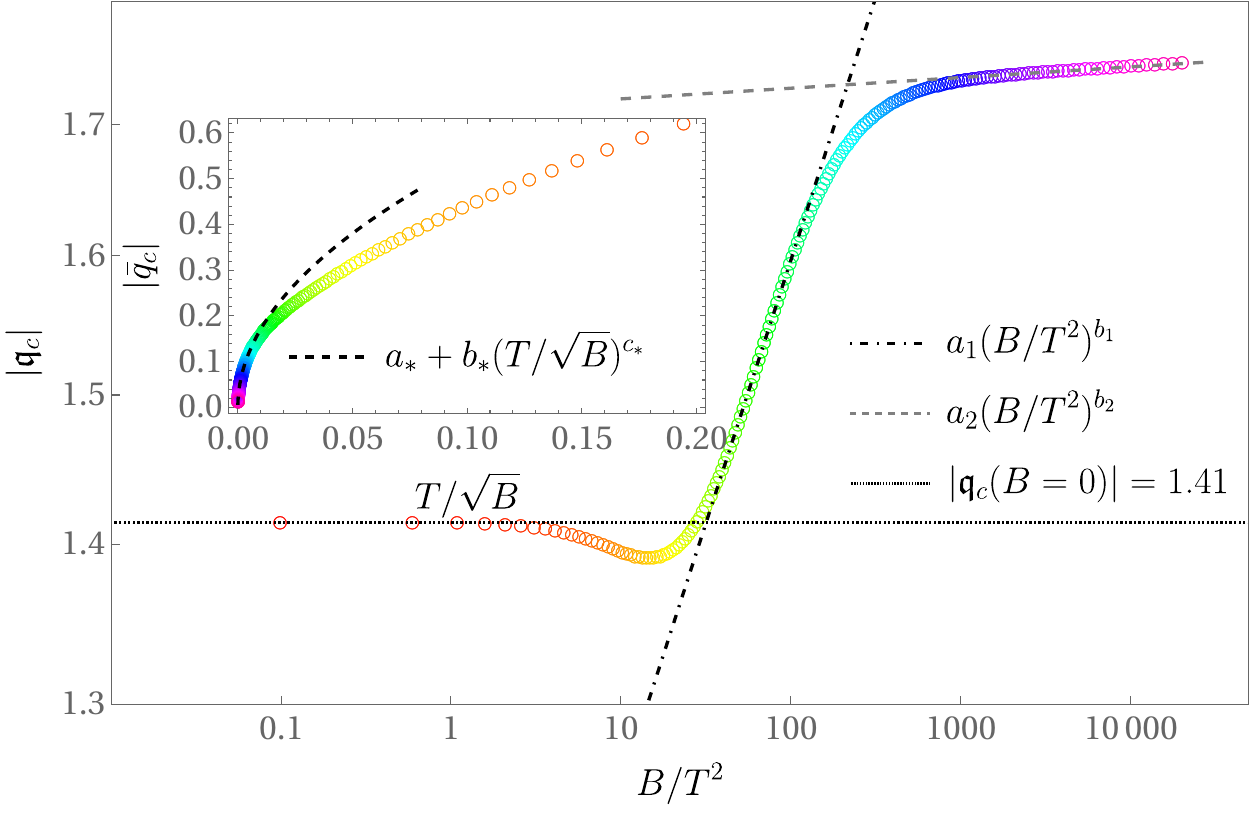}
    \caption{\textit{Magnetic Convergence:} The radius of convergence of the linearized hydrodynamic expansion of $\mathcal{N}=4$ SYM minimally coupled to a global $U(1)$ gauge theory subjected to strong external fields. The horizontal dashed line indicates the value of the convergence radius for $B=0$ as originally found in~\cite{Grozdanov:2019uhi}. The fit parameters are recorded in table~\ref{tab:fitParaConvergence}. The inset graphic displays the same data but normalized by the magnetic field rather then the temperature. In both images the colors match those in Fig.~\ref{fig:1pfamily} and Fig.~\ref{fig:Complex_q_plane}
\label{fig:magnetic_Convergence}}
\end{figure}
\begin{figure}[htbp]
    \centering
    \includegraphics[width=0.47\textwidth ]{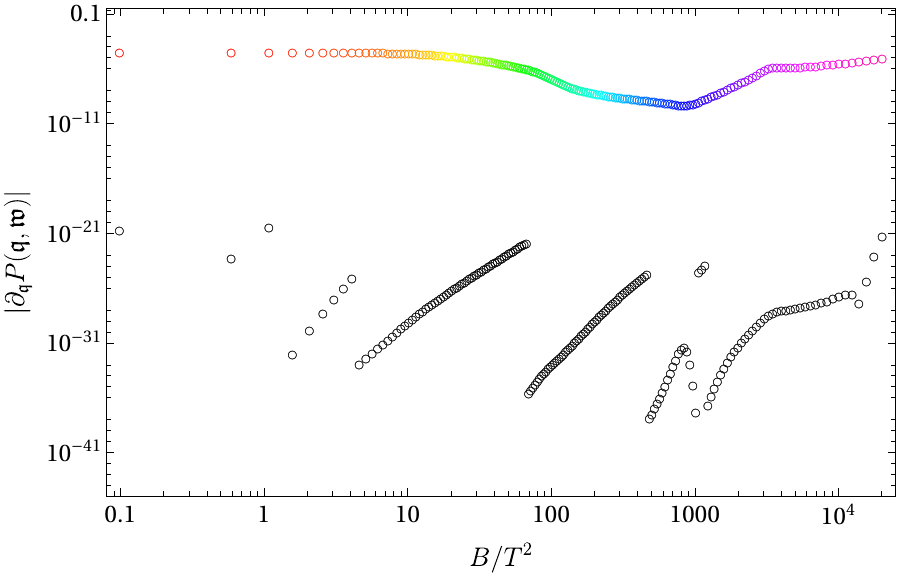}
    \caption{\textit{Momentum derivative of the spectral curve:} The absolute value of $\partial_\q P_{Reg}(\q_c,\w_c)$ as a function of the dimensionless magnetic field $B/T^2$ is displayed as colored open circles, where the colors match those in Fig.~\ref{fig:1pfamily} and Fig.~\ref{fig:Complex_q_plane}. The black open circles display the regulated value of  the curve $P_{Reg}(\q,\w)$. In both cases the values are regulated by $P(1,1;B/T^2)$, that is the value of the spectral curve evaluated at $(\q,\w)=(1,1)$ for each value of the dimensionless magnetic field value $B/T^2$.
\label{fig:Complex_q_Der}}
\end{figure}

Moving further away from the hydro regime, both the momentum and frequency at the critical points are pushed from their original location closer to the imaginary axis. This is displayed in Fig.~\ref{fig:Complex_q_plane} by the color, red at the lowest values of $B/T^2$, purple at largest value of the magnetic field. This behavior is similar to the behavior of quasinormal modes found in~\cite{Ammon:2017ded}. There, the authors noted that for large magnetic fields the modes were found to be drawn towards the imaginary axis. Given the smooth behavior it may be expected that as $\lim_{B/T^2\rightarrow \infty} Re(\q) =0$, although this has not been confirmed. The origin of this behavior is very likely related to the results of the previous sections which display that critical points, at least of spin $s=0$ operators, in a $2d$ CFT occur at purely imaginary values of the frequency and momentum.

As discussed in section~\ref{sec:Det_Method} the critical point which is closest to the origin encodes the bound on the radius of convergence of the hydrodynamic dispersion relations. Shown in Fig.~\ref{fig:magnetic_Convergence} is absolute value of the critical momentum i.e. the radius of convergence of the hydrodynamic dispersion relations of $\mathcal{N}=4$ SYM minimally coupled to $U(1)$ gauge theory subjected to external magnetic fields. Interestingly even when $B/T^2=1$ the radius of convergence does not change substantially, with the relative difference between $B/T^2\approx 1/10$ and $B/T^2\approx 1$ remaining on the order of $10^{-4}$. This is quite surprising, $B/T^2\approx 1$ is well beyond the regime where hydrodynamics should be applicable.

\begin{figure*}[htbp]
    \centering
    \includegraphics[width=0.47\textwidth]{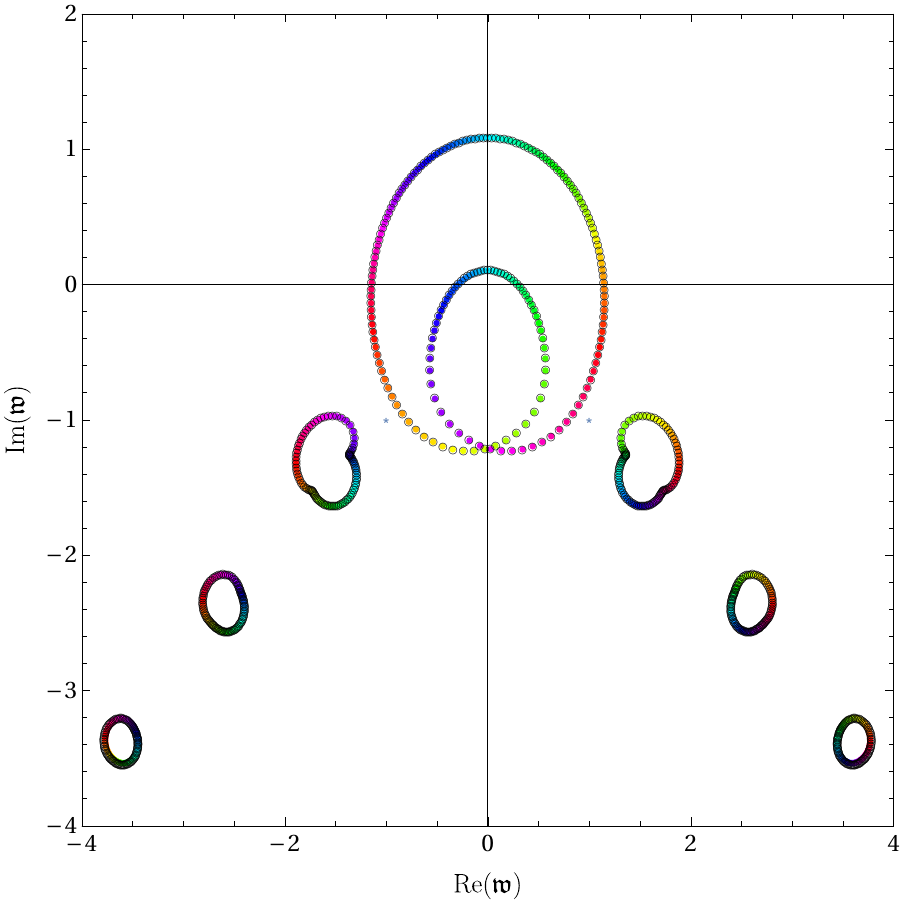} \hfill
    \includegraphics[width=0.47\textwidth]{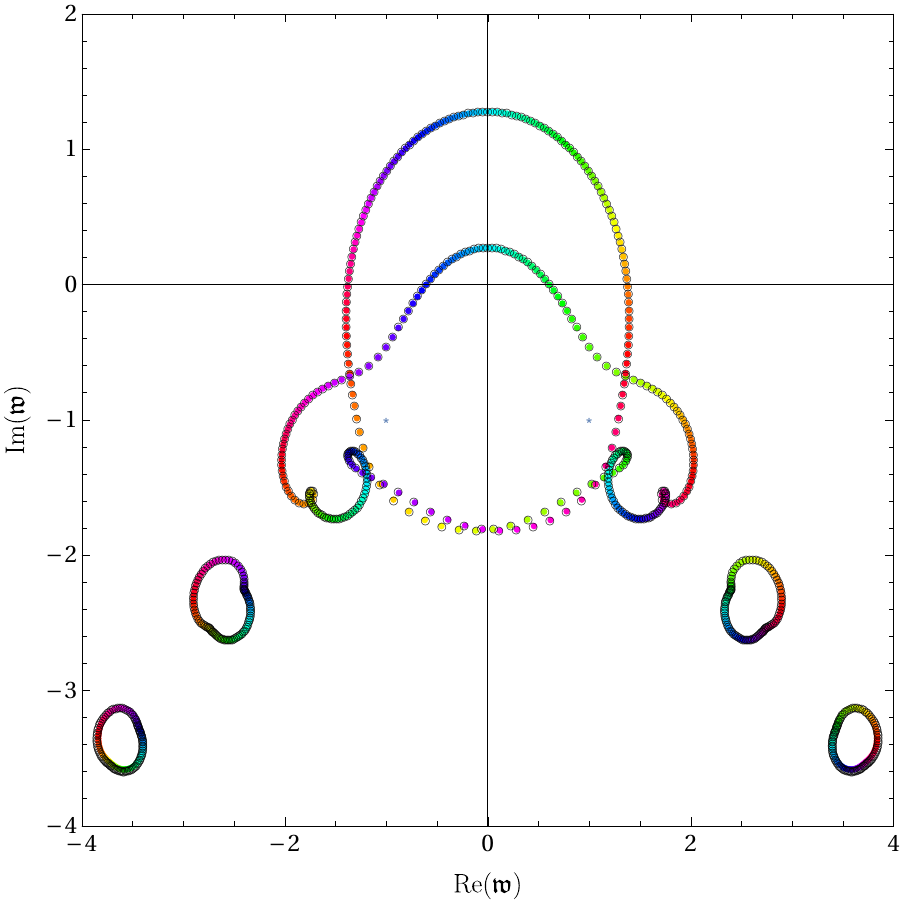}
    \caption{\textit{Monodromy I}: The complex frequency plane in which we display the monodromy $\w(|x_i|^2 e^{i\phi})$ with $i=1,2$ for $|x_1|<|\q_c|<|x_2|$ and $\phi\in[0,2\pi]$. In both images these curves are displayed compared to the case of the Schwarzchild black brane ($B/T^2=0$) displayed as empty black dots while the curves for the magnetic black brane (with $B/T^2\approx \mathbf{1}$ ) are displayed as a spectrum of colors. The color varies from \textcolor{red}{red} ($\phi=0$), through \textcolor{cyan}{cyan} ($\phi=\pi$), to \textcolor{magenta}{magenta} ($\phi=2\pi$). In each case $|x_1|=|\q_c|-1/10$ and $|x_2|=|\q_c|+1/10$. It is easy to see visually that even for magnetic field strengths on the order of $B/T^2\approx 1$, the location of the critical point remains roughly the same.
 \textit{Left:} The magnitude of the momentum is slightly below the value of the critical momentum. \textit{Right:} The magnitude of the momentum is slightly above the value of the critical momentum.
    \label{fig:monodromy_1}}
\end{figure*}

\begin{figure*}[hbtp]
    \centering
    \includegraphics[width=0.47\textwidth]{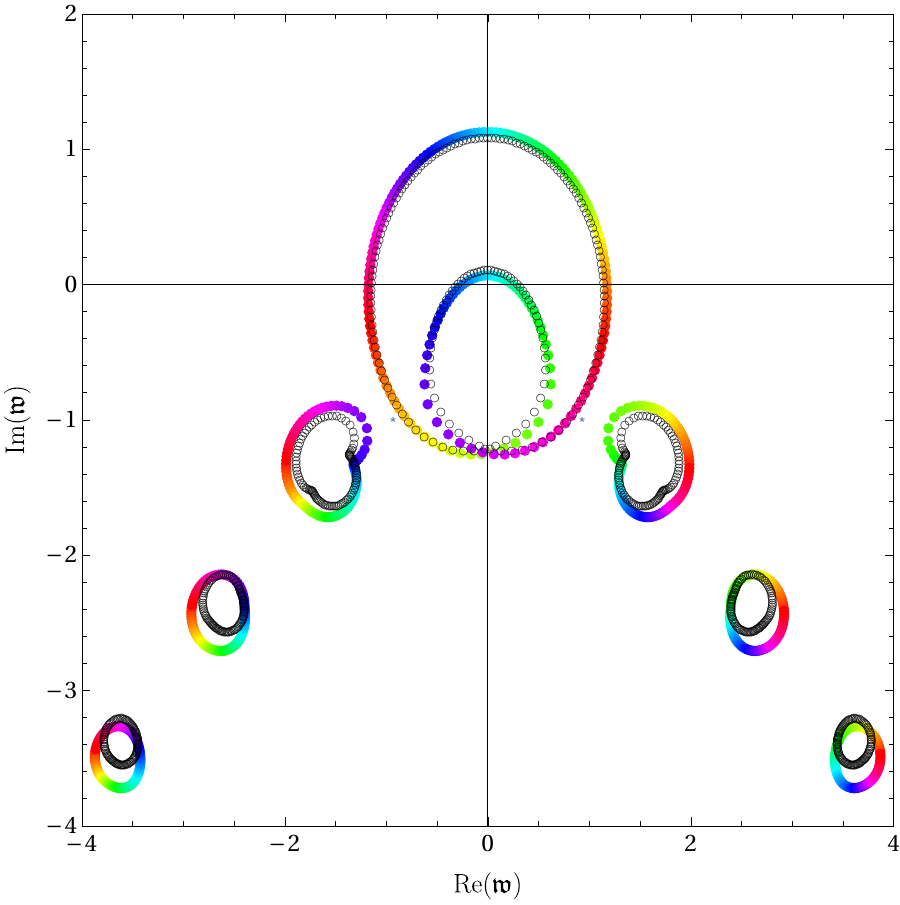} \hfill
    \includegraphics[width=0.47\textwidth]{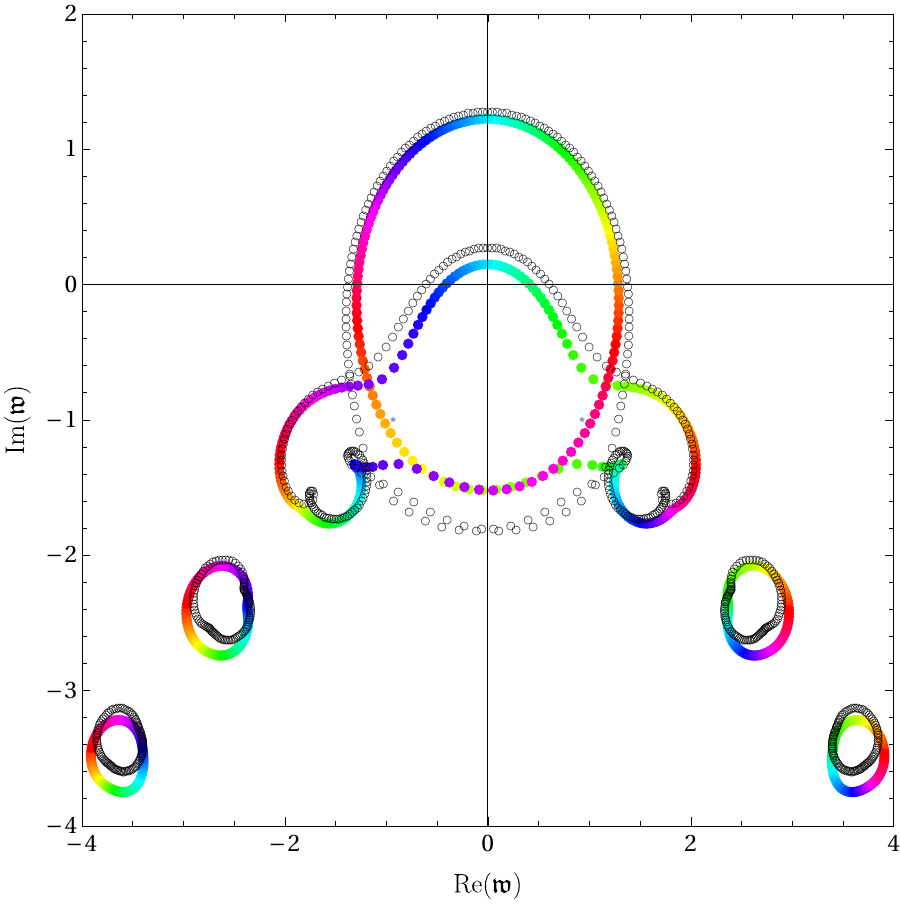}
    \caption{\textit{Monodromy II}: The complex frequency plane in which we display the monodromy $\w(|x_i|^2 e^{i\phi})$ with $i=1,2$ for $|x_1|<|\q_c|<|x_2|$ and $\phi\in[0,2\pi]$. In both images these curves are displayed compared to the case of the Schwarzchild black brane ($B/T^2=0$) displayed as empty black dots while the curves for the magnetic black brane (with $B/T^2\approx \mathbf{10}$ ) are displayed as a spectrum of colors. The color varies from \textcolor{red}{red} ($\phi=0$), through \textcolor{cyan}{cyan} ($\phi=\pi$), to \textcolor{magenta}{magenta} ($\phi=2\pi$). In each case $|x_1|=|\q_c|-1/10$ and $|x_2|=|\q_c|+1/10$.
 \textit{Left:} The magnitude of the momentum is slightly below the value of the critical momentum. \textit{Right:} The magnitude of the momentum is slightly above the value of the critical momentum.
    \label{fig:monodromy_19}}
\end{figure*}

Before moving to larger field values, we should first pause and make sure that the modes we compute are indeed critical points, rather then singular points. In Fig.~\ref{fig:Complex_q_Der} we display $|\partial_\q P|$ as a function of the dimensionless magnetic field strength. One can see this remains above our numerical value for zero throughout the whole range of $B/T^2$ considered in this work. Although, this is a numerical determination from a regulated function, hence we should be very cautious with this plot. When possible, we should seek to verify this in another way, and for magnetic fields up to approximately $B/T^2\approx 100$ we have been able to confirm this visually. This is displayed Fig.~\ref{fig:monodromy_1} (at $B/T^2\approx 1$) and Fig.~\ref{fig:monodromy_19} (at $B/T^2\approx 10$) which display the monodromy of the frequency. For both figures the left plot displays a magnitude of the momentum smaller the critical momentum and the right plot shows a magnitude of the momentum larger the critical momentum. A star is used to denote $\w(\q_c)=\w_c$. Here, the color no longer indicates the magnetic field, but rather the value of the phase angle as described in the plot caption. In both cases we can visually see distinct sets of curves in the left images, but after the magnitude of the momentum becomes larger then the critical momentum the three sets of modes closest to the line $\text{Im}(\w)=0$ join and become one large connected set. Once this happens the once three separate modes interchange positions as $\phi$ runs from $0$ to $2\pi$.

\

Having built our confidence that we indeed are looking at critical points of the curve we return to Fig.~\ref{fig:magnetic_Convergence} and continue investing its behavior at larger magnetic field values. We find that increasing the magnetic field leads to a decrease in the radius of convergence. Between the values of $1\lesssim B/T^2\lesssim 28$ there is a dip in the convergence radius with a minimum value of $|\q_c|= 1.391$ at $B/T^2\approx 14.56$. Interestingly this is a local minimum and the radius of convergence begins to grow after reaching it. It is at $B/T^2\approx 28$ where the radius of convergence returns to the same value as a neutral plasma at finite temperature. The radius of convergence then sees an enhancement due to the magnetic field growing in size roughly like $a_1(B/T^2)^{b_1}$ for $a_1$ and $b_1$ given in table~\ref{tab:fitParaConvergence}.
\begin{table}[htbp] \caption{Fit parameters measuring the growth of the critical momentum $\q_c$ \label{tab:fitParaConvergence}}
    \centering
    \begin{ruledtabular}
    \begin{tabular}{l|l|l }
    Fit function $|\q_c|$ & $a_i$ & $b_i$  \\
    \hline
     $ a_1 (B/T^2)^{b_1}$ &0.980 & 0.105 \\
    $ a_2 (B/T^2)^{b_2}$  & 1.711  & 0.002
    \end{tabular}
    \end{ruledtabular}
\end{table}
As $B/T^2$ begins to approach large values ($B/T^2> 1000$) the growth rate of the convergence radius slows considerably. The goodness of fit between a polynomial function and a constant function deviate only slightly and a polynomial fit of the form $|q_c|=a_2 (B/T^2)^{b_2}$ returns an exponent of $b_2=0.00204$. While we have been unable to work in the strict $B/T^2\rightarrow \infty$ limit, this suggests the radius of convergence of a hydrodynamic expansion in this limit may be a constant number, given approximately by $|q_c(B/T^2\rightarrow \infty)|\approx 1.745$. We might therefore expect that there exists a hydrodynamic description of large field regime of $\mathcal{N}=4$ SYM theory. Analytic solutions to the Einstein field equations exist~\cite{DHoker:2009mmn} in this regime and it remains an open question to apply these techniques to the geometry at large field values. Finally we note that, as seen in Fig.~\ref{fig:1pfamily}, our background continues to become closer and closer BTZ$_{2+1}\times \mathbb{R}^2$. It is reasonable to suspect that eventually $\partial_\q P=0$ and the critical points transition to singular points. Here the observed level-crossing in the QNM spectrum transitions to level-touching and the dispersion relations become analytic functions. Hence the radius of convergence in this limit is not finite, but rather the location of pole collision nearest to the origin in the complex plane remains at a finite value. We will comment more on this before the end of the section.

It is important to consider this as the limit that $B\gg T$, not the limit in which $T\rightarrow 0$. Considering this as a zero temperature limit requires looking more carefully at the geometry. A naive investigation with the data at hand requires a different normalization, $(\bar{q},\bar{\omega})= T/\sqrt{B}(\q,\w)$. Rearranging the data one finds that $\bar{q}$ is fit well by~\footnote{The value of $b_*$ is inconsequential for the argument here, however for completeness the fit was done with a fit function of the form $a_*+b_* y^{c_*}$ for data $\bar{q}(T/\sqrt{B})$. Insisting that at zero temperature the radius of convergence is greater or equal to zero gives $a=0, b_*=1.675, c_*=0.495$. } $\bar{q}\approx b_* (T/\sqrt{B})^{1/2}$ and hence as $T\rightarrow 0$ so does the critical momentum suggesting a zero radius of convergence hydrodynamic expansion at zero temperature. The data normalized in this way is displayed as the inset graphic in Fig.~\ref{fig:magnetic_Convergence}. However hydrodynamic expansions at zero temperature require a more careful treatment, see for instance~\cite{Davison:2013bxa}, hence the naive statements above should be taken lightly. A description of solutions at zero temperature do exist~\cite{DHoker:2009mmn} and it remains an open question to apply these techniques to this zero temperature case.

Before closing this section, we take a moment to try to understand a little better what is happening to the pole collisions for large fields. As discussed throughout this work, the geometry we have focused on can be considered as an renormalization group flow, triggered by the magnetic field, from a $3+1$ CFT to a $1+1$ CFT. The dimensional reduction, from $d$ to $d-2$ dimensional physics brought on a magnetic field is referred to as magnetic catalysis~\cite{Gusynin:1994re,Gusynin:1994va,Gusynin:1994xp}, and happens quite generally (see for instance~\cite{Miransky:2015ava} for a review of such phenomena in systems ranging from QCD to Dirac semimetals). In the current system, the dimensionally reduced 1+1 dimensional field theory may be described as a Luttinger liquid~\cite{Giamarchi:2003sfp} (or sometimes a Luttinger-Tomonaga liquid) as clearly demonstrated in~\cite{DHoker:2010xwl,DHoker:2011ehc} via the direct calculation of the renormalization group flow of correlation functions in the magnetic black brane geometry considered here~\footnote{See also~\cite{C.Biagini_2001} where it is demonstrated in general that strong magnetic fields induce Luttinger liquid phases in bulk metals}. As summarized in~\cite{DHoker:2009mmn,DHoker:2010xwl} in the free field limit, the eigenfunctions are Landau levels and hence are localized in the $x_{1,2}$ directions but have arbitrary momentum along the magnetic field lines. For low energy only the lowest Landau level is occupied and hence only the momentum along the magnetic field is the remaining quantum number. Furthermore the lowest Landau level energy is $E=0$, while for bosons it is $E\sim \sqrt{B}$ hence for energies less then $\sqrt{B}$ only the lowest Landau level participates and the theory is a $1+1$ CFT of fermionic excitations.

Having settled our picture of the low energy physics at large magnetic field, we are must now contend with the hydrodynamic description of such CFTs. Fortunately, hydrodynamic behavior of $1+1$ CFTs have been discussed in~\cite{Haehl:2018izb}, in terms of the modern interpretation of fluid dynamics, that of a non-linear sigma model of mappings between a fixed reference, or world volume, spacetime and a dynamical, or physical spacetime. Here, the hydrodynamic theory can be considered as a field theory of the soft modes associated with holomorphic and anti-holomorphic reparameterizations. Roughly speaking, these arise from considering conformal transformations $\delta z=\epsilon(z)$ generated by the currents $J(z)=\epsilon(z)T(z)$ (where $T$ is the holomorphic sector of the energy momentum tensor). This symmetry transformation becomes a gapless mode if one considers a generic dependence on the anti-holomorphic parameter i.e. $\epsilon(z,\bar{z})$. The Lorentzian propagator, modulo a choice of pole prescription, can be written as
\begin{align}
    G(\omega,k)&=\braket{\epsilon(\omega,k)\epsilon(-\omega,-k)}\nonumber \\
    &=-\frac{24\pi}{c}\frac{1}{\omega (\omega^2+1)(\omega-k)}
\end{align}
From the poles we can obtain the spectral function $P=\omega (\omega^2+1)(\omega-k)$ and obtain that $(\omega,k)=(0,0)$ and $(\omega,k)=(\pm i,\pm i)$ correspond to level-touching points. Hence there are no critical points which limit the radius of convergence of the hydrodynamic expansion of the dispersion relations. This is easy to see since the dispersion relation is analytic, and precisely given as a dissipationless sound mode,  $\omega=k$, hence this will converge for all values of $k$ except $k\rightarrow \infty$.

However, we are then confronted with a discrepancy, the value of the mode extracted from the numerical analysis at large fields does not seem to be approaching this value. How do we reconcile this? To understand this difference we need to recall the structure of the metric given in Eq.(\ref{eq:BTZ_IR_Metric}. As in~\cite{DHoker:2009mmn}, the geometry we consider is an interpolation between AdS$_5$ and $BTZ\times \mathbb{R}^2$ (or a compact space $V^2$ i.e. $BTZ\times V^2$ if we want the dual theory to be a genuine CFT with central charge proportional to the volume of the space $V^2$). Looking at the metric in Eq.~\ref{eq:BTZ_IR_Metric} we can see that metric of the CFT would be $\exd s^2_{IR-CFT}=-3\exd t^2+3\exd x_3^2$. Therefore, we must perform a rescaling of the coordinates used to obtain the Greens function $G(\omega,k)$, $(t,x_3)$, if we wish to compare to the expected IR result given by $(\omega,k)=(\pm i,\pm i)$. Naively we would have that $i (-\omega t + k x_3)= i (-\omega \sqrt{3} t'+k \sqrt{3} x_3')$. Hence we should expect $(\omega',k')=\sqrt{3}(\pm i,\pm i)$. Comparing to the value found by the numerics, $(|\omega|,|k|) = (1.759,1.749)$ it appears as though we are approaching the expected IR value with a percent difference of $1.55\%$ and $1.01\%$ for the frequency and momentum respectively. However, it is worth pausing to note that we have not completed the renormalization group flow. Our numerics go only to $B/T^2\approx O(10^5)$, for which the metric functions already are beginning to develop strong gradients near the horizon (this is noted also in~\cite{Ammon:2016szz,Cartwright:2021hpv}) and are beginning to breakdown. Hence it is unclear if these modes indeed asymptote to the expected IR values, or, if these modes move off to new locations in the complex plane and the expected pole collision of the IR CFT is the result of two different modes (not currently captured by our numerics) interacting. To understand which of these scenarios is the case would require new techniques to increase the accuracy of the numerics and go beyond the magnetic field strengths considered in this work.

\section{Discussion}
In this work we considered the radius of convergence of dispersion relations in $\mathcal{N}=4$ SYM theory minimally coupled to a global $U(1)$ gauge field. The effective description took the form of external strong field hydrodynamics. We constructed a numerical representation of the hydrodynamic spectral curve and used simple root finding techniques to obtain the location of critical points. To provide some reassurance that we were indeed investigating critical points we computed the QNM spectrum to visually inspect the monodromy of the frequency under phase rotations in the complex momentum plane. In addition, we have also introduced a simple new method to discern between level-touching and level-crossing points which has yet to appear in the literature. We demonstrated its validity by 1) appealing to the definition of r-fold points given in~\cite{walker:1950alg}, 2) explicitly illustrating its use in a simple example and 3) by using it to explain the level-touching phenomena of BTZ black holes that had been seen in previous works. With the critical points we obtained, we constructed the radius of convergence of hydrodynamic dispersion relations in this theory and demonstrated that they remain finite for $B/T^2$ values far exceeding the naive statement for the regime of hydrodynamic applicability (i.e. $B/T^2\ll 1$). In fact, rather then further limiting the validity of the dispersion relations, increasing the strength of the magnetic field ultimately leads to an increase in the radius of convergence of the series.

In the intermediary regime, $1\lesssim B/T^2\lesssim 28$ there is a dip in the convergence radius with a minimum value of $|\q_c|= 1.391$ at $B/T^2\approx 14.56$. Although it is unclear why this occurs, it is interesting to note it occurs when the magnetic field strength is on the same order of magnitude as the temperature scale.

In the large field regime the growth of the radius of convergence slows considerably and it appears that the radius of convergence (or perhaps more cautiously, the location of the pole collision nearest to the origin) approaches a finite value $|q_c(B/T^2\rightarrow \infty)|\approx 1.748$. As indicated by Fig.~\ref{fig:1pfamily}, it is in this limit that our geometry truly appears to be BTZ$_{2+1}\times\mathbb{R}^2$ and hence it may be reasonable to suspect analytic dispersion relations such as those derived in effective field theory descriptions of the energy-momentum tensor in AdS$_3$/CFT$_2$~\cite{Haehl:2018izb,Datta:2019jeo}. A naive comparison to the soft modes associated with holomorphic and anti-holomorphic reparmeterizations appears to show our analysis agrees to roughly the percent level. However, this is precisely the location where we have the least trust in the in numerics. At very large magnetic field it becomes challenging to obtain the QNM spectrum and hence we have no visual means of checking the monodromy of the curve. While we have trust in the measure that $\partial_\q P\neq 0$ to ensure a critical point, the numerical estimation of this measure requires working with a regulated value. We must therefore be cautious with the results obtained at large magnetic field. Clearly new techniques must be devised to work with large magnetic fields. A simple place to start would be to use the analytic solutions to the field equations in this regime~\cite{DHoker:2009mmn} and repeat this analysis perturbatively in $1/B$ corrections to the IR geometry. This potential transition to analytic dispersion relations driven by extreme magnetic field strengths is an interesting phenomena. We hope to better understand this in the future.

Furthermore, a naive fit of the appropriately normalized data predicts a hydrodynamic expansion with vanishing radius of convergence at zero temperature. It would be an interesting task to more careful construct a hydrodynamic expansion in this limit and check this statement.

We again emphasize that this study was conducted for a theory with a global $U(1)$ gauge field in the dual field theory, leading to strong \textit{external} field hydrodynamics. It would be rewarding to extend this study to the case of local $U(1)$ gauge fields, hence these gauge fields would be dynamically determined and the effective description would be true MHD. A general proof of finite radii of convergence of the dispersion relations predicted by MHD would be a challenging endeavour. However, it is possible to potentially provide an example of how moving to dynamical gauge fields changes the behavior of these series for holographic theories. The authors of~\cite{Grozdanov:2017kyl} demonstrate how one can induce renomalization group flows by double trace operators and arrive at theories whose long-wavelength, small frequency effective theory is MHD. Extending the results of~\cite{Grozdanov:2019uhi} and of this current work to the case of a MHD description of $\mathcal{N}=4$ SYM theory is certainly of interest.

As discussed in~\cite{Grozdanov:2017kyl} the choice of renormalization scale is related to the choice of the renormalized electromagnetic coupling $e_r$. The choice in this work roughly sets the UV scale to be that of the magnetic field. Hence, this is a reasonable choice when studying the strong field hydrodynamic expansion as done in this work. However, the choice made is not the only choice, and as demonstrated in~\cite{Grozdanov:2017kyl}, the choice made of the renormalization point scale, and hence the electromagnetic coupling, can have a strong effect. In some instances this can lead to an infinite $U(1)$ coupling or an imaginary coupling and hence issues such as instabilities and superluminal propagation. This will be a challenging analysis, and, it will be necessary to work in a framework that is valid in both the weak field and strong field regime. Since our goal was an investigation of strong field external hydrodynamics we worked with both a framework and renomalization point convenient for such a study. However, a thorough investigation of the dependence of the radius of convergence on the renormalization point would be a highly interesting endeavour.

In addition, we note that the effect of the chiral anomaly is not taken into account here since $\bar{\gamma}=0$. In a follow up paper we intend to investigate its effects on the critical points of the hydrodynamic dispersion relations. As a preliminary statement, including the chiral anomaly requires leaving $\bar{\gamma}\neq 0$ and enhancing the ansatz in Eq. (\ref{eq:metric_ansatz}) and Eq. (\ref{eq:Gauge_Ansatz}) to a more general ansatz (see for instance~\cite{DHoker:2009ixq,Ammon:2017ded,Cartwright:2021hpv}) which includes a temporal component of the gauge field, providing the field theory with a chemical potential $\mu$. Doing so one finds the $x_3$ component of the current contains the chiral magnetic effect e.\ g.\ $\braket{J_3}\sim \bar{\gamma} \mu B$. For small values of the chemical potential $\mu/T=1/10$, while $\bar{\gamma}$ takes on its supersymmetric value, one can repeat the work of section II and III and find as expected that the deviations of the nearest critical point from the origin are small.  However, unlike the magnetic field, the chemical potential introduces new effects on the radius of convergence of the dispersion relations including piece-wise continuous behavior (see~\cite{Abbasi:2020ykq,Jansen:2020hfd} for the case without magnetic field). The same holds with the Chern-Simons coupling and magnetic field present. The details associated with these calculations will be discussed in a forthcoming publication.

\acknowledgments
The author is grateful to the organizers of the ECT$^*$ program ``Holographic Perspectives on Chiral Transport" for an invitation to present early results from this work. The author thanks ECT* for support at the program ``Holographic Perspectives on Chiral Transport" during which part of this research had been developed and completed. The author is also grateful to the participants of this excellent workshop for comments and discussion. The author is also grateful to Sebastian Grieninger for discussion and specifically to Sa\u{s}o Grozdanov for comments and questions on an early draft of this work. This work is supported by the Netherlands Organisation for Scientific Research (NWO) under the VICI grant VI.C.202.104.

\bibliographystyle{apsrev4-1}
\bibliography{bib}

\end{document}